\documentclass[aps, prx, twocolumn, superscriptaddress, amsmath, amssymb, longbibliography, floatfix]{revtex4-1}

\usepackage[final]{graphicx}
\usepackage{hyperref}
\usepackage{subfigure}
\usepackage{enumerate}
\graphicspath{{figures/}}
\usepackage{color}

\usepackage{xr}
\usepackage{dcolumn}% Align table columns on decimal point

\begin{document}

\title{Site Mixing for Engineering Magnetic Topological Insulators}

\author{Yaohua Liu}\email[]{liuyh@ornl.gov}
\affiliation{Neutron Scattering Division, Oak Ridge National Laboratory, Oak Ridge, TN 37831, USA}

\author{Lin-Lin Wang}
\affiliation{Division of Materials Science and Engineering, Ames Laboratory, Ames, IA 50011, USA}

\author{Qiang Zheng}
\affiliation{Department of Materials Science and Engineering, University of Tennessee, Knoxville, TN 37996, USA}

\author{Zengle Huang}
\affiliation{Department of Physics and Astronomy, Rutgers University, Piscataway, NJ 08854, USA}

\author{Xiaoping Wang}
\affiliation{Neutron Scattering Division, Oak Ridge National Laboratory, Oak Ridge, TN 37831, USA}

\author{Miaofang Chi}
\affiliation{Materials Science and Technology Division, Oak Ridge National Laboratory, Oak Ridge, TN 37831, USA}

\author{Yan Wu}
\affiliation{Neutron Scattering Division, Oak Ridge National Laboratory, Oak Ridge, TN 37831, USA}

\author{Bryan C. Chakoumakos}
\affiliation{Neutron Scattering Division, Oak Ridge National Laboratory, Oak Ridge, TN 37831, USA}

\author{Michael A. McGuire}
\affiliation{Materials Science and Technology Division, Oak Ridge National Laboratory, Oak Ridge, TN 37831, USA}

\author{Brian C. Sales}
\affiliation{Materials Science and Technology Division, Oak Ridge National Laboratory, Oak Ridge, TN 37831, USA}

\author{Weida Wu}
\affiliation{Department of Physics and Astronomy, Rutgers University, Piscataway, NJ 08854, USA}

\author{Jiaqiang~Yan}\email[]{yanj@ornl.gov}
\affiliation{Materials Science and Technology Division, Oak Ridge National Laboratory, Oak Ridge, TN 37831, USA}
\affiliation{Department of Materials Science and Engineering, University of Tennessee, Knoxville, TN 37996, USA}

\date{\today}

\begin{abstract}

The van der Waals compound, MnBi$_2$Te$_4$, is the first intrinsic magnetic topological insulator, providing a materials platform for exploring exotic quantum phenomena such as the axion insulator state and the quantum anomalous Hall effect. However, intrinsic structural imperfections lead to bulk conductivity, and the roles of magnetic defects are still unknown. With higher concentrations of same types of magnetic defects, the isostructural compound MnSb$_2$Te$_4$ is a better model system for a systematic investigation of the connections among magnetic, topology and lattice defects. In this work, the impact of antisite defects on the magnetism and electronic structure is studied in MnSb$_2$Te$_4$. Mn-Sb site mixing leads to complex magnetic structures and tunes the interlayer magnetic coupling between antiferromagnetic and ferromagnetic. The detailed nonstoichiometry and site-mixing of MnSb$_2$Te$_4$ crystals depend on the growth parameters, which can lead to  $\approx$40\% of Mn sites occupied by Sb and $\approx$15\% of Sb sites by Mn in as-grown crystals. Single crystal neutron diffraction and electron microscopy studies show nearly random distribution of the antisite defects. Band structure calculations suggest that the Mn-Sb site-mixing favors a FM interlayer coupling, consistent with experimental observation, but is detrimental to the band inversion required for a nontrivial topology. Our results suggest a long range magnetic order of Mn ions sitting on Bi sites in MnBi$_2$Te$_4$. The effects of site mixing should be considered in all layered heterostructures that consist of alternating magnetic and topological layers, including the entire family of MnTe(Bi$_2$Te$_3$)$_n$, its Sb analogs and their solid solution. 

 \end{abstract}

 \maketitle

\section{Introduction}
%Structural imperfections can give rise to new and extraordinary properties that do not exist in ideal crystals.
Structural imperfection in crystalline materials can have profound effects on the electronic, magnetic, optical, mechanical, and thermal properties. For topological insulators, the surface state is protected against nonmagnetic lattice defects due to spin-momentum locking. Magnetic dopants have been purposely introduced to topological insulators to induce magnetism, which breaks time reversal symmetry. This approach led to the first experimental observation of the quantum anomalous Hall effect (QAHE) in Cr$_{0.15}$(Bi$_{0.1}$Sb$_{0.9}$)$_{1.85}$Te$_3$ films~\cite{chang2013experimental}.  This complex chemical formula highlights the importance of identification and fine manipulation of lattice defects in realizing the exotic properties of topological materials. Because of the random distribution of Cr in the Bi layer, however, the QAHE was only observed at extremely low temperatures of 30\,mK, even though the onset of magnetic ordering occurs at $\approx$15\,K~\cite{chang2013experimental}.

\begin{figure*} \centering \includegraphics [width = 0.9\textwidth] {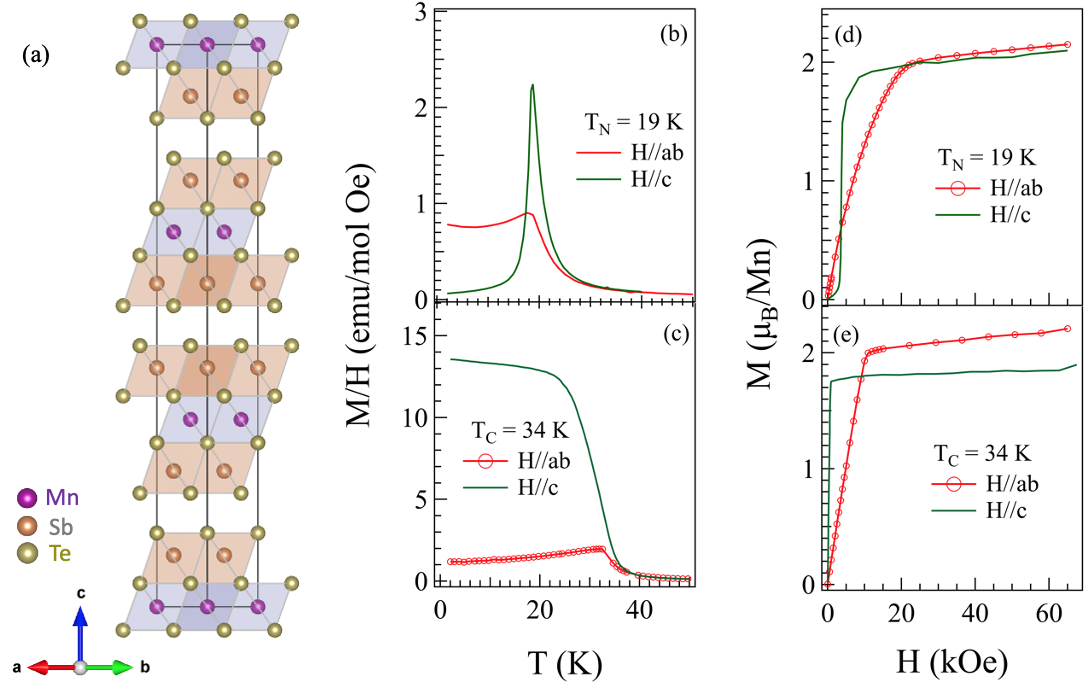}
\caption{(color online) Different growth temperatures result in distinct magnetic properties of MnSb$_2$Te$_4$ single crystals. (a) Ideal crystal structure of MnSb$_2$Te$_4$, isostructural to MnBi$_2$Te$_4$. (b),(c) Temperature dependence of magnetic susceptibility measured in a field-cooled mode with an applied magnetic field of 1\,kOe. (d),(e) Field dependence of magnetization at 2\,K. The applied magnetic field is either along ($H||c$) or perpendicular to ($H||ab$) the crystallographic \textit{c}-axis. The crystal with T$_N$\,=\,19\,K was grown at 620~$^\circ$C, while the one with T$_C$\,=\,34\,K at 640~$^\circ$C.}
\label{fig:Mag}
\end{figure*}

MnBi$_2$Te$_4$, a natural heterostructure of magnetic (MnTe) and topological (Bi$_2$Te$_3$) ingredients, in principle should be an ideal platform for observing the QAHE and axion behavior since it is a stoichiometric compound instead of a random alloy. However, MnBi$_2$Te$_4$, like any crystalline compound, will have defects. As shown in Fig.\,\ref{fig:Mag}(a), MnBi$_2$Te$_4$ consists of septuple layers of Te-Bi-Te-Mn-Te-Bi-Te stacked along the crystallographic \textit{c}-axis~\cite{lee2013crystal}. Below $T_N$=25\,K, the high spin Mn$^{2+}$ ($d^5$, $S=5/2$) ions order ferromagentically in each septuple layer which then couple antiferromagnetically with neighbouring layers forming the so-called A-type antiferromagnetic (AFM) structure~\cite{otrokov2019prediction,yan2019crystal}. Therefore, in addition to those lattice defects observed in Bi$_2$Te$_3$~\cite{scanlon2012controlling,cava2013crystal,ando2013topological,hashibon2011first,zhou2012controlling,chuang2018anti}, magnetic defects are expected in MnBi$_2$Te$_4$. The question then becomes what type and what concentration of defects can be tolerated to preserve the entangled quantum ground state? How do these defects impact the magnetism and the observation of the QAHE? These questions are very general and apply to any  material proposed for observation of the QAHE.

Experimental investigations~\cite{yuan2020electronic,zhu2020investigating,yan2019crystal,zeugner2019chemical,li2020antiferromagnetic} reported antisite disorders of Mn$_{Bi}'$ (Mn at the Bi site with one negative charge following Kroger-Vink notation), Bi$_{Mn}^.$ (Bi at Mn site with a positive charge), Bi$_{Te}'$ and likely Mn vacancies in MnBi$_2$Te$_4$ crystals or films. Density functional theory (DFT) calculations~\cite{du2020tuning} suggest the lattice mismatch between MnTe and Bi$_2$Te$_3$ sheets facilitates the formation of Mn-Bi site mixing. These lattice defects lead to the bulk conductivity and affect the spin wave spectra~\cite{li2020competing,li2020two}. However, it is unknown whether the Mn ions at Bi site order magnetically,  how the magnetic defects affect the magnetism and whether they are related to the diverse surface states resolved by angle-resolved photoemission (ARPES) studies~\cite{otrokov2019prediction, zeugner2019chemical,lee2019spin,swatek2020gapless,vidal2019surface,chen2019topological,li2019dirac,hao2019gapless}. Furthermore, ample experimental evidence suggests these magnetic defects also exist in other  MnTe(Bi$_2$Te$_3$)$_n$ members with n$>$1~\cite{souchay2019layered,wu2019natural,yan2020type,vidal2019topological,hu2019realization,tian2019magnetic,liang2020mapping,wu2020toward,wu2020distinct}.  Therefore, it is critical to understand the formation, distribution, consequences, and manipulation of the magnetic defects in this family of compounds.

MnSb$_2$Te$_4$ is a model system for such a systematic investigation of the connections among magnetism, topology, and lattice defects. MnSb$_2$Te$_4$ is isostructural to MnBi$_2$Te$_4$, and these two compounds have the same types of defects\cite{du2020tuning}. However, the concentration of lattice defects in MnBi$_2$Te$_4$ is only a few percent~\cite{yan2019crystal}, which is difficult to systematically study. In contrast, MnSb$_2$Te$_4$ tends to have higher concentrations of magnetic defects because the difference of ionic size and electronegativity between Mn and Sb is much smaller than that between Mn and Bi in MnBi$_2$Te$_4$. From crystal growth point of view, MnSb$_2$Te$_4$ crystals can be grown easily in a wide temperature range which makes it possible fine tune of the concentrations of lattice defects and hence the magnetism by varying the growth temperatures. The growth of MnBi$_2$Te$_4$ has been quite challenging and the crystallization occurs in a narrow temperature window of $\approx$10~$^\circ$C~\cite{yan2019crystal, zeugner2019chemical}. Finally, while MnSb$_2$Te$_4$ single crystals are reported to order antiferromagnetically~\cite{yan2019evolution}, polycrystalline samples are reported to have ferrimagnetic septuple layers that are coupled ferromagnetically due to the presence of Mn at Sb site~\cite{murakami2019realization}. It is unknown whether this complex magnetic structure is inherent to the polycrystalline sample or the discrepancy between single crystal and polycrystalline samples results from different amount of magnetic defects and signals their essential roles. Of particular interest is the origin of the ferromagnetic (FM) inter-septuple-layer coupling, which might be employed to fine tune the magnetism of MnBi$_2$Te$_4$ for the observation of QAHE at elevated temperatures.

In this work, we, therefore, perform a systematic study of the magnetic defects in MnSb$_2$Te$_4$ single crystals with different magnetic ground states and tunable magnetic ordering temperatures. Our results show that the randomly distributed Mn-Sb antisite defects favor a FM interlayer coupling but  are detrimental to the band inversion necessary for a non-trivial band topology. Our results imply that Mn$_{Bi}'$ ions in MnBi$_2$Te$_4$ order ferromagnetically below T$_N$=25\,K but they are antiferromagentically coupled to Mn$_{Mn}^\times$ ions in each septuple layer. Partial substitution of Bi by magnetic ions in MnBi$_2$Te$_4$ might be a valid approach toward a FM inter-septuple-layer coupling and quantum Hall effect at zero magnetic field.

\section{Experimental details}
MnSb$_2$Te$_4$ single crystals were grown out of an Sb-Te flux~\cite{yan2019crystal}. After homogenizing the starting materials at 900~$^\circ$C for overnight, the mixture was furnace cooled to a certain temperature in between 620~$^\circ$C and 640~$^\circ$C and kept at the selected temperature for two weeks. The fixed-temperature growth was employed to obtain uniform crystals because the lattice defects are sensitive to the growth temperatures and starting compositions. The crystals were then separated from Sb-Te flux by decanting. In this work, we focus on the crystals grown at three different temperatures of 620~$^\circ$C, 630~$^\circ$C, and 640~$^\circ$C. The growth temperature determines the nonstoichiometry and site mixing and thus the magnetic order. We also tried crystal growths starting with different ratio of MnTe:Sb$_2$Te$_3$, which also affects the nonstoichiometry and site mixing in MnSb$_2$Te$_4$ and enables a fine tuning of the magnetic ordering temperature up to 50\,K (Fig.~\ref{fig:Tunable} in Supporting Information). 

Elemental analysis on cleaved surfaces was performed using either the energy dispersive (EDS) or the wavelength dispersive (WDS) spectroscopy techniques. The EDS measurement was performed using a Hitachi TM-3000 tabletop electron microscope equipped with a Bruker Quantax 70 energy dispersive x-ray system. The WDS measurement was performed using a JEOL JXA-8200X electron microprobe analyzer instrument equipped with five crystal-focusing spectrometers for waveength dispersive x-ray spectroscopy.  Magnetic properties were measured with a Quantum Design (QD) Magnetic Property Measurement System in the temperature range 2.0\,K$\leq T \leq$\,300\,K. The temperature and field dependent electrical resistivity data were collected using a QD Physical Property Measurement System.

High resolution single crystal neutron diffraction data were collected at 300~K at TOPAZ at the Spallation Neutron Source (SNS) to determine the average nuclear structures. The integrated Bragg intensities are obtained using the 3D ellipsoidal Q-space integration method and are corrected for background using MANTID software~\cite{michels2016expanding}. Data reduction including Lorentz and absorption corrections as well as spectrum, detector efficiency, data scaling, and normalization are carried out with the ANVRED3 program~\cite{schultz1984single}. The structure models were analyzed with the SHELX prgoram~\cite{sheldrick2008short}. Low temperature single crystal neutron diffraction experiments were carried out at beamline CORELLI~\cite{ye2018implementation} at SNS and the Four-Circle Diffractometer (HB3A) at the High Flux Isotope Reactor (HFIR) to determine the average magnetic structure.  Possible magnetic structures were investigated by the representation analysis using the SARAh program~\cite{wills2000new}, as well as the magnetic space group approach, where the maximal magnetic space groups were obtained from the MAXMAGN program~\cite{perez2015symmetry}. Nuclear and magnetic structure refinements were carried out with the FullProf Suite program~\cite{rodriguez1993recent}. Additionally, the temperature dependence of the magnetic order and potential diffuse scattering in all three samples were investigated at CORELLI.

Scanning transmission electron microscopy (STEM) observations were performed on a Nion UltraSTEM100, equipped with a cold field-emission gun and a corrector of third- and fifth-order aberrations, operated at the accelerating voltage of 100~kV.  About $2\times2~\mu$m$^2$ thin specimens of MnSb$_{2}$Te$_{4}$ were prepared by focused-ion-beam (FIB), and subsequently by ion milling at a low voltage of 1.5\,kV and liquid nitrogen temperature for 20\,min. High-angle annular dark-field (HAADF)-STEM images were collected with a probe convergence angle of 30\,mrad and an inner collection angle of 86\,mrad. Scanning tunneling microscopy (STM) measurements were performed at 48\,K in an Omicron LT-STM with base pressure $1\times10^{-10}$~mbar. Electrochemically etched tungsten tips were characterized on clean Au(111) surface before STM experiments. Single crystals of MnSb$_{2}$Te$_{4}$ were cleaved in situ at room temperature and immediately inserted into a cold STM head.

Band structure with spin-orbit coupling (SOC) in density functional theory~\cite{hohenberg1964inhomogeneous,kohn1965self} (DFT) have been calculated with PBE~\cite{perdew1996generalized} exchange-correlation functional, a plane-wave basis set and projected augmented wave method as implemented in VASP~\cite{kresse1996efficient,kresse1996efficiency}. To account for the half-filled strongly localized Mn 3d orbitals, a Hubbard-like U~\cite{dudarev1998electron} value of 3.0\,eV is used. The ($2\times2\times2$) MnSb$_2$Te$_4$ rhombohedral unit cell is sampled with a Monkhorst-Pack~\cite{monkhorst1976special} (6$\times$6$\times$3) k-point mesh including the $\Gamma$ point and a kinetic energy cutoff of 270\,eV. For band structure calculations, the experimental lattice parameters have been used with atoms fixed in their bulk positions. The supercell has also been fully relaxed in PBEsol~\cite{perdew2008restoring} exchange correlation functional with U for total energy difference.

\begin{figure*} \centering \includegraphics [width = 0.9\textwidth] {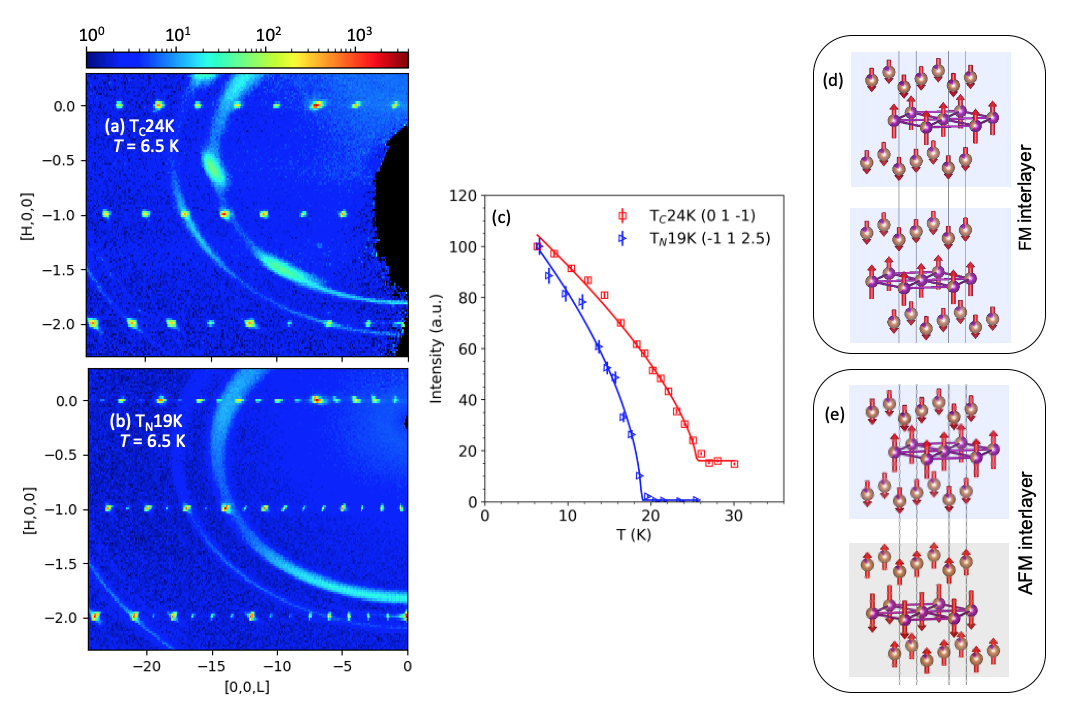}
\caption{(color online) Distinct interlayer coupling in different MnSb$_2$Te$_4$ single crystals resolved by single crystal neutron diffraction. (a),(b) Neutron scattering pattern in the [H~1~L] plane of the T$_C$24K and T$_N$19K, respectively. The data were collected at 6.5~K, where the sample T$_N$19K shows Bragg peaks at half-$L$ positions but T$_C$24K does not.  (c) Temperature dependence of the normalized intensities of selected magnetic Bragg reflections for T$_N$19K and T$_C$24K. The solid lines are guides to the eye. (d),(e) The schematic diagram of the magnetic structures of T$_C$24K and T$_N$19K, respectively. Both have the same intra-septuple-layer ferrimagnetic order, but the FM inter-septuple-layer coupling for T$_C$24K is in contrast to the AFM one for T$_N$19K. The magnetic space groups for them are $R\bar{3}m^\prime$ (no. 166.101) and \textit{R}$_I$-3\textit{c} (no. 167.108), respectively. The sites connected by the purple triangular lattice represent the MnTe sheets while the other sites represent Mn mixed into the Sb$_2$Te$_3$ sheets.}
\label{fig:neutron}
\end{figure*}

\section{Results}
\subsection{Distinct magnetic properties}
Figures\,\ref{fig:Mag}(b)-\ref{fig:Mag}(e) show two distinct types of magnetic properties of our MnSb$_2$Te$_4$ crystals. An antiferromagnetic(AFM)-like anisotropic magnetic susceptibility (see Fig.\,\ref{fig:Mag}(b)), $\chi$, for the crystals grown at 620~$^\circ$C is in contrast to a FM-like temperature dependence (see Fig.\,\ref{fig:Mag}(c)) for those grown at 640~$^\circ$C.  The magnetic susceptibility for both batches shows a similar temperature dependence when the external magnetic field is applied perpendicular to the crystallographic \textit{c}-axis. However, when the magnetic field is applied along the \textit{c}-axis, $\chi_c$ in Fig.\,\ref{fig:Mag}(b) decreases sharply when cooling below 19\,K; while $\chi_c$ in Fig.\,\ref{fig:Mag}(c) shows a rapid increase upon cooling below the magnetic ordering temperature. The former is similar to that observed for MnBi$_2$Te$_4$ suggesting an AFM interlayer coupling~\cite{yuan2020electronic,zhu2020investigating,yan2019crystal,zeugner2019chemical,li2020antiferromagnetic,yan2019evolution}. The latter suggests a FM interlayer coupling, as previously observed in polycrystalline samples~\cite{murakami2019realization}.  This is also supported by the field dependence of magnetization shown in Figs.\,\ref{fig:Mag}(d) and~\ref{fig:Mag}(e).  A well defined spin flip transition is observed in Fig.\,\ref{fig:Mag}(d) when the field is applied along the \textit{c}-axis. This spin flip transition is absent in Fig.\,\ref{fig:Mag}(e), also suggesting a FM interaction along the \textit{c}-axis. When the magnetic field is applied perpendicular to the \textit{c}-axis, a similar field dependence is observed for both crystals; however, the magnetic field at which the magnetization tends to saturate decreases from about 25\,kOe in Fig.\,\ref{fig:Mag}(d) to 10\,kOe in Fig.\,\ref{fig:Mag}(e). The saturation moment obtained by extrapolating $M$($H$) curve to $H=0$ is 1.95(2) and 1.98(2)~$\mu_B$/Mn, respectively. For both crystals, the anisotropic field and temperature dependencies of magnetization suggest the ordered magnetic moments are along the \textit{c}-axis.

The magnetic measurements suggest that the sign of the interlayer interaction is tunable by the growth temperature. The crystals grown below 620~$^\circ$C show the AFM-like $\chi$($T$), as in Fig.\,\ref{fig:Mag}(b); and those grown above 625~$^\circ$C show a FM-like behavior, as in Fig.\,\ref{fig:Mag}(c). While all AFM-like crystals order magnetically at the same $T_N$, the magnetic ordering temperature of FM-like crystals can be tuned in the temperature range 24\,K-50\,K by controlling the growth parameters as shown in Fig.~\ref{fig:Tunable} in Supporting Information. As described later, the magnetic ordering temperature and saturation moment depend on the nonstoichiometry and distribution of Mn. In the following text, we refer to those three crystals that are thoroughly investigated as T$_N$19K, T$_C$24K, and T$_C$34K, respectively, with T$_N$ indicating AFM-interlayer coupling, T$_C$ for FM-interlayer coupling, and the number as the magnetic ordering temperature.

\subsection{Site mixing and magnetic structure from neutron single crystal diffraction}
The anisotropic magnetic properties suggest that the interlayer coupling can be either FM or AFM although in both cases the ordered moment is aligned along the \textit{c}-axis. To understand the different magnetic behaviors and determine the detailed magnetic structures, we performed single crystal neutron diffraction experiments in a wide temperature range using multiple diffractometers. High-resolution diffraction data collected at 300\,K at beamline TOPAZ confirmed that the average structure can be well described by the space group $R\bar{3}m$ for both types of crystals (Fig.~\ref{fig:Rn3m} in Supporting Information), similar to previous reports on powder~\cite{murakami2019realization} or single crystal samples~\cite{yan2019evolution,zhou2020topological}. Neutron diffraction is sensitive to the Mn-Sb mixing because of the high scattering contrast between Mn and Sb nuclei. Refinement of neutron diffraction data show significant Mn-Sb mixing at both Mn and Sb sites for all samples with about 13-16\% Mn$_{Sb}'$ and 32-41\% Sb$_{Mn}^.$. The chemical formulas are (Mn$_{0.588(2)}$Sb$_{0.412(2)}$)(Sb$_{0.871(2)}$Mn$_{0.129(2)}$)$_2$Te$_4$, (Mn$_{0.635(2)}$Sb$_{0.365(2)}$)(Sb$_{0.850(2)}$Mn$_{0.150(2)}$)$_2$Te$_4$and (Mn$_{0.674(4)}$Sb$_{0.326(4)}$)(Sb$_{0.842(3)}$Mn$_{0.158(3)}$)$_2$Te$_4$ for T$_N$19K, T$_C$24K, T$_C$34K, respectively (Table S1-S4 in Supporting Information). The refined elemental ratio agrees with that obtained from elemental analyses as summarized in Table S6 in Supporting Information. Interestingly, with increasing growth temperatures, MnSb$_2$Te$_4$ crystals contain more Mn ions on both Mn and Sb sites. This influences the competing magnetic interactions hence the sign of inter-septuple-layer coupling and the magnetic ordering temperature.

Low temperature diffraction data were collected at CORELLI to investigate the magnetic structure and potential diffuse scattering. Datasets for refinement were collected for T$_N$19K and T$_C$24K at both 30\,K and 6.5\,K. The diffraction patterns collected above magnetic ordering temperature are similar for both crystals (e.g. Fig.~S2(a) in Supporting Information).  However, they are quite different in the magnetically ordered state at 6.5\,K. The diffraction pattern in Fig.\,\ref{fig:neutron}(a) for T$_C$24K is similar to that collected at 30\,K but the intensity of some peaks is enhanced as highlighted by the white arrows (Fig.\,S2 in Supporting Information), suggesting a magnetic wavevector of $k_{FM}~=~(0~0~0)$. T$_N$19K (see Fig.\,\ref{fig:neutron}(b) shows a completely new set of peaks at the half-L positions below T$_N$, suggesting a magnetic wavevector of $k_{AFM}~=~(0~0~1.5)$. Fig.\,\ref{fig:neutron}(c) shows the temperature dependence of intensities of selected Bragg peaks which confirms the magnetic ordering wavevectors as well as the magnetic ordering temperatures as found from magnetic and transport measurements.

\begin{figure*} \centering \includegraphics [width = \textwidth] {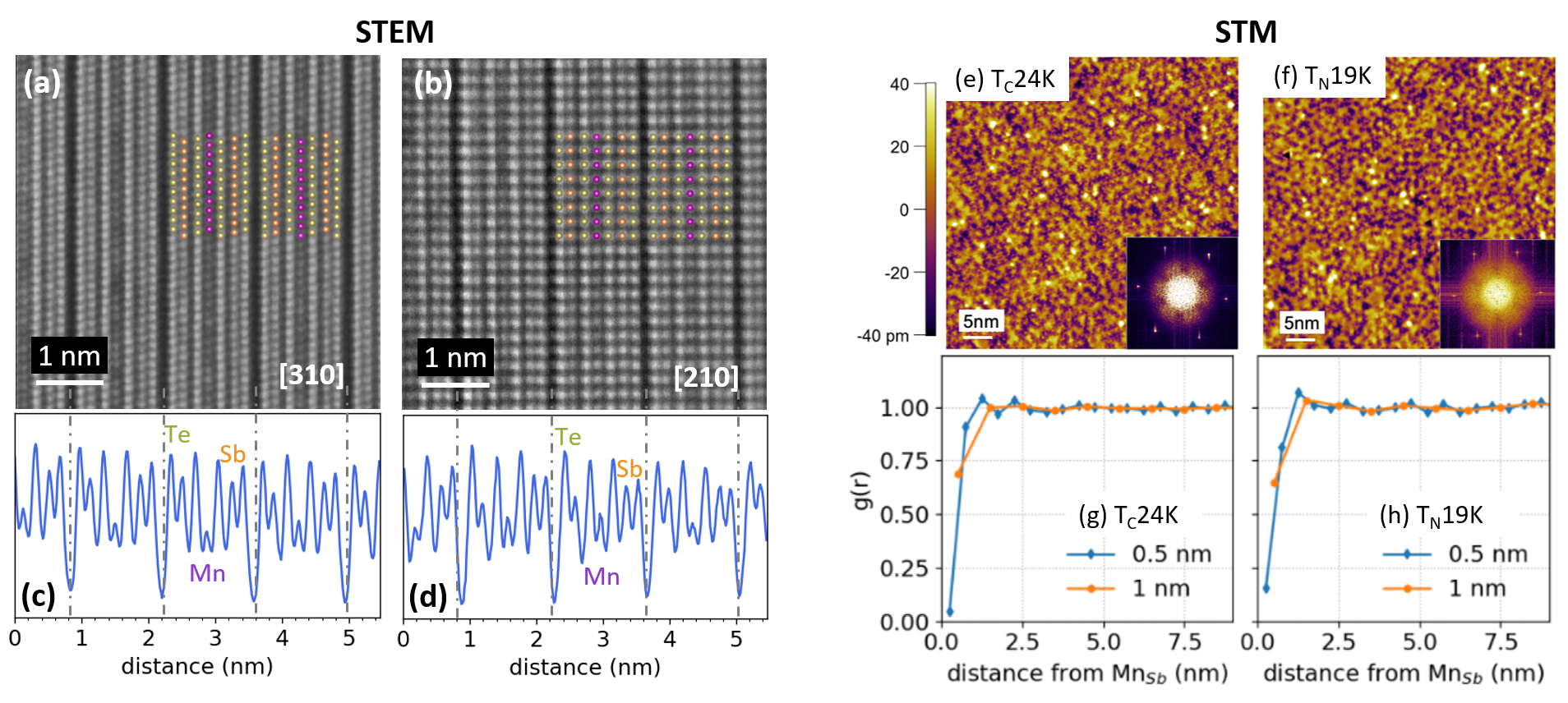}
\caption{(color online) Presence and random distribution of Mn-Sb antisite defects in both MnSb$_2$Te$_4$ crystals from STEM and STM. (a),(b) HAADF-STEM images along [310] for T$_N$19K (a) and along [210] for T$_C$24K (b). No stacking fault was observed. (c),(d) Sum intensity profile perpendicular to the septuple layers. (e) STM topographic image of T$_C$24K (tunneling condition: -0.4V, 100 pA). (f) STM topographic image of T$_N$19K (tunneling condition: -0.4V, 60 pA). The insets in (e) and (f) are the Fast Fourier transform (FFT) of the topographic images. (g),(h) Radial pair distribution function of Mn$_{Sb}'$ antisites. Radial pair distribution function $g$($r$) measures the normalized density of Mn$_{Sb}'$ in the distance of r away from a reference Mn$_{Sb}'$.  Two bin sizes, 0.5\,nm and 1\,nm are used to calculate $g$($r$).}
\label{fig:STEMSTM}
\end{figure*}

The refinement on the CORELLI data shows the same amount of Mn-Sb site mixing as determined from TOPAZ measurements (Table\,S5 in Supporting Information). The absence of structured diffuse scattering patterns suggests that the antisite defects do not induce significant local lattice distortions, and they are either randomly distributed or have a correlation length shorter than the probing limiting $\approx$~1\,nm. There is no magnetic Bragg scattering contribution for (0 0 L)-type peaks for both T$_N$19K and T$_C$24K (data not shown), suggesting that the ordered moment is along the $c$-axis. This is consistent with the anisotropic magnetic properties presented above.  Figures~\ref{fig:neutron}(c) and~\ref{fig:neutron}(d) show the magnetic structures determined from the low temperature diffraction datasets for T$_C$24K and T$_N$19K, respectively. As shown in Fig.\,S3 in Supporting Information, the magnetic order of Mn$_{Sb}'$ ions has to be considered leading to the ferrimagnetic septuple layer for both magnetic structures. Both Mn$_{Mn}^\times$ and Mn$_{Sb}'$ ions carry significant ordered moments and they couple antiferromagnetically in each septuple layer. At 6.5\,K, the ordered moment of $\approx$4.0$\mu_B$/Mn for Mn$_{Mn}^\times$ is slightly larger than $\approx$3.5$\mu_B$/Mn for Mn$_{Sb}'$. A distinct difference between the magnetic structures shown in Figs.~\ref{fig:neutron}(d) and~\ref{fig:neutron}(e) is the sign of the inter-septuple-layer coupling. A FM interlayer coupling shown in Fig.~\ref{fig:neutron}(d) is in contrast to an AFM one in Fig.~\ref{fig:neutron}(e).

\begin{figure*} \centering \includegraphics [width = 0.9\textwidth] {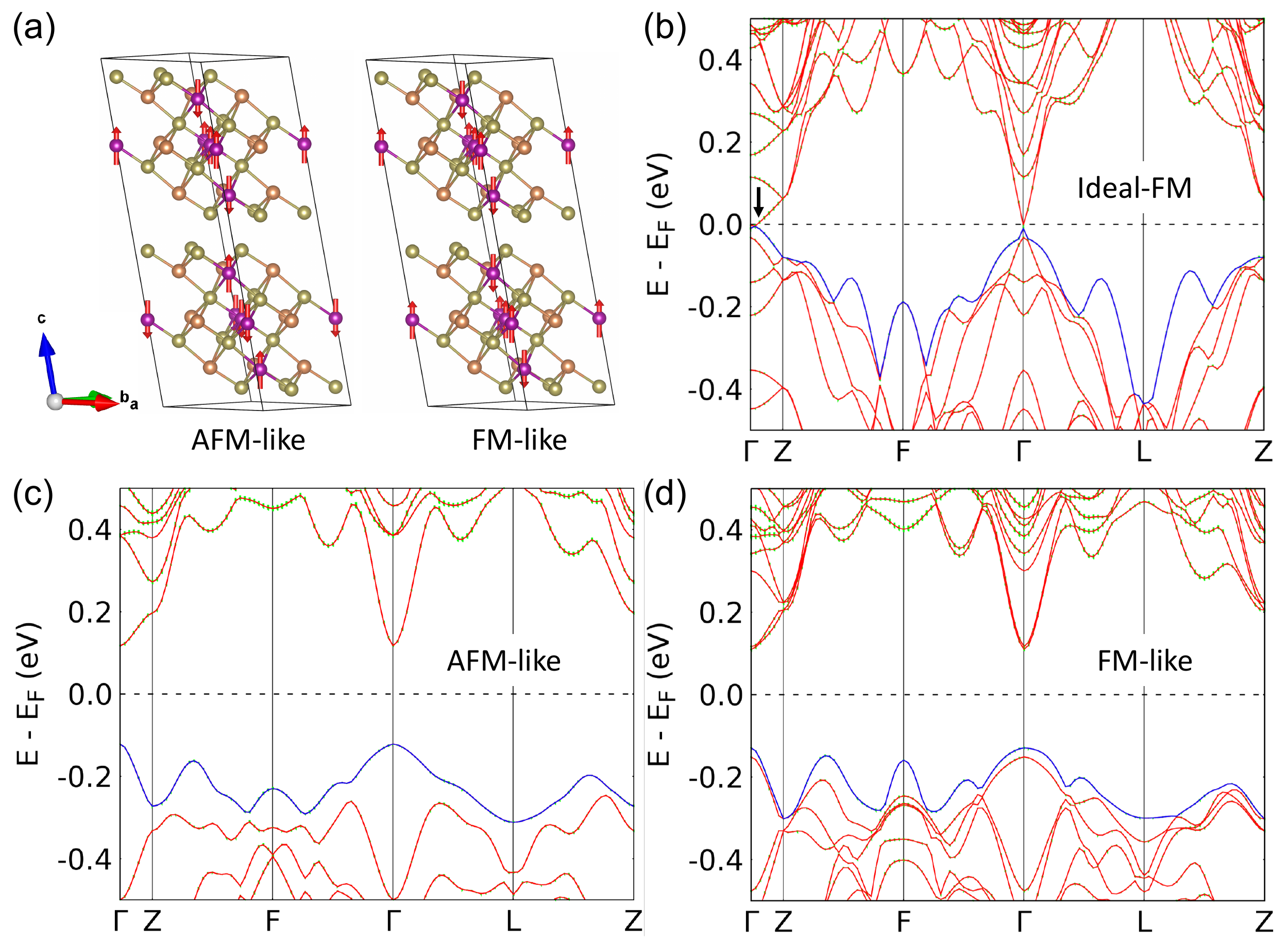}
\caption{(color online) Mn-Sb site mixing favors FM interlayer coupling but is detrimental to the band inversion from DFT.(a) (2$\times$2$\times$2) supercells of rhombohedral MnSb$_2$Te$_4$ in the most stable configuration with the Mn-Sb site-mixing ratio of 50/50 in the central layer and 25/75 in the Sb layer, showing AFM or FM interlayer couplings while keeping the
inversion symmetry. This configuration (labelled as AS3) and the other inequivalent ones are listed in Table S7 and shown in Fig.~S8 of the Supporting Information. PBE+U+SOC band structures of (b) the ideal-FM without site-mixing, where the black arrow indicates the Weyl point in ideal-FM along $\Gamma$-Z due to band inversion. (c),(d) AFM-like and FM-like configurations with Mn-Sb site mixing. The top valence band is in blue and the green shadow indicates the projection on the $p$ orbital of Sb at Mn site.}
\label{fig:Band}
\end{figure*}

\subsection{Presence and random distribution of antisite defects from STEM and STM}
We further investigated the presence and distribution of Mn-Sb antisites and looked for other potential lattice defects in T$_C$24K and T$_N$19K using STEM and STM. Figures\,\ref{fig:STEMSTM}(a) and~\ref{fig:STEMSTM}(b) show the HAADF-STEM images along the layer directions for T$_C$24K and T$_N$19K, respectively, showing the Te-Sb-Te-Mn-Te-Sb-Te septuple layers stacking along the [001] direction. Since the HAADF-STEM image intensity is roughly proportional to Z$^2$ (Z is the atomic number), the Mn (Z = 25) atomic columns should show weaker intensity, while the Te (Z = 51) and Sb (Z = 52) atomic columns should display similar stronger intensities. However, the intensities of some Sb columns are weaker compared to the Te columns (in Figs.\,\ref{fig:STEMSTM}(c) and~\ref{fig:STEMSTM}(d), indicating Mn$_{Sb}'$ antisite defects. A careful inspection (Fig.~\ref{fig:STEMProfile} in Supporting Information) of the intensity variation of Mn and Te columns in (001) planes found a larger intensity variation among Mn atomic columns, indicating the partial substitution of Mn by heavier atoms. The STEM images for T$_C$24K and T$_N$19K look the same. No stacking faults or other lattice defects were identified.

The distribution of Mn at Sb site is further investigated by the STM measurement. The typical STM topographic images of T$_C$24K and T$_N$19K are shown in Figs.\,\ref{fig:STEMSTM}(e) and~\ref{fig:STEMSTM}(f) respectively.  In the Fast Fourier transform in the insets, the sharp Bragg peaks indicate the atomic resolution of the images. The atomic resolution allows identification of Mn$_{Sb}'$ and determination of its density on the surface. Like Mn$_{Bi}'$ in MnBi$_2$Te$_4$~\cite{yan2019crystal,yuan2020electronic}, Mn$_{Sb}'$ in MnSb$_2$Te$_4$ exists as a dark triangle in STM topography due to the depression in the three nearest-neighbor Te atoms in the topmost plane right above the Mn$_{Sb}'$. The density of Mn$_{Sb}'$ is estimated to be $\approx$10.6\% in T$_C$24K, slightly larger than $\approx$8.4\% in T$_N$19K. The ratio is about 1.26, consistent with our neutron scattering results. To explore the distribution of Mn$_{Sb}'$, the radial distribution function analysis is performed after locating each Mn$_{Sb}'$. The radial distribution function $g$($r$) describes how the density of defects varies as a function of distance, by calculating the normalized density of defects within a distance between $r-\frac{1}{2}dr$ and $r+\frac{1}{2}dr$ from a reference defect, where $dr$ is the binsize. Here two binsizes of 0.5\,nm and 1\,nm are used. As shown in Figs.\,\ref{fig:STEMSTM}(f) and~\ref{fig:STEMSTM}(g), $g$($r$) is smaller than 1 at a short distance of 0.5\,nm, and fluctuates around 1 at larger distances, indicating some Mn$_{Sb}'$-Mn$_{Sb}'$ repulsion with a short correlation scale. In contrast, the clustering of defects would result in $g$($r$) much larger than 1 at short distances, which is absent in our analysis. A careful investigation of our STM images suggests $\approx$0.3\% of Sb$_{Te}'$ defects in both T$_C$24K and T$_N$19K crystals.

\subsection{Electronic structure calculations}

To understand the effects of site mixing on the magnetic order and electronic structure, we performed DFT calculations. We first constructed a (2$\times$2$\times$2) rhombohedral supercell at stoichiometry with the Mn-Sb site-mixing ratio of 50/50 in the central Mn layer and 25/75 in each of the Sb layers. With the same alternating Mn and Sb rows in the central Mn layer in the (2x2x2) supercell (see Figs.~\ref{fig:Config}(b)-~\ref{fig:Config}(f) in Supporting Information), all the possible five inequivalent configurations of a Mn mixed in each of the two Sb layers, or 25/75, have been considered. Figure 4(a) shows the most stable configuration in the (2$\times$2$\times$2) supercell (labelled as AS3 in Table S7 and Fig.~\ref{fig:Config} in Supporting Information) with the complex magnetic structure. Additionally we also considered two configurations in the larger (3$\times$3$\times$2) supercell with the Mn-Sb site-mixing ratio of 56/44 in the central layer and 22/78 in the Sb layer, where the central Mn layer has a mixture of Mn and Sb in each row, unlike the alternating rows in (2$\times$2$\times$2). As listed in Table S7, after full relaxation in PEBsol+U~\cite{perdew2008restoring,dudarev1998electron}, a FM interlayer coupling is preferred for all the antisite configurations, in a distinct contrast to the ideal case without site-mixing. Such a preference for all the configurations also holds when SOC is included during the full relaxation. As a reference, for the ideal MnSb$_2$Te$_4$ without site-mixing, the A-type AFM structure is 3.48 meV/f.u. more stable than the FM order for PBEsol+U with relaxation. The (2$\times$2$\times$2)-AS3 complex magnetic configuration in Fig.~\ref{fig:Band}(a) with an FM interlayer coupling is 0.31~meV/f.u. more stable than that with an AFM interlayer coupling. Although in the above supercells the variation  for composition and configuration is limited compared to the experimental compositions, these total energy differences confirm that the magnetic coupling between the septuple layers changes the preference from AFM to FM with more Mn at Sb site, in qualitative agreement with experimental observation.

We further  studied the effects of Mn-Sb site mixing on the electronic structure. First, our calculations show that the ideal MnSb$_2$Te$_4$ with an A-type AFM order has a small gap~\cite{murakami2019realization,zhou2020topological,chen2019intrinsic} without band inversion and is topologically trivial. In contrast, the ideal MnSb$_2$Te$_4$ with a FM order is topologically non-trivial supported by the Weyl point (black arrow) along $\Gamma$-Z in Fig.\,\ref{fig:Band}(b) due to the band inversion between Sb-derived conduction and Te-derived valence band. Stabilizing FM in MnSb$_2$Te$_4$ is promising to realize magnetic Weyl points, similar to EuCd$_2$As$_2$~\cite{wang2019single,jo2020manipulating}. These results are consistent with previous reports~\cite{murakami2019realization,zhou2020topological,chen2019intrinsic}. We then introduced Mn-Sb site mixing as discussed above and the electronic band structures of the most stable configuration in the (2$\times$2$\times$2) supercell with AFM and FM interlayer couplings are shown in Figs.\,\ref{fig:Band}(c) and~\ref{fig:Band}(d), respectively. As shown in Fig.\,\ref{fig:Band}(c), each band is still doubly degenerated because the supercell retains inversion symmetry and the effective time-reversal symmetry~\cite{mong2010antiferromagnetic}. The combined time-reversal operation and half-translation along the \textit{c}-axis is the same as in the ideal A-type AFM configuration. But the band gap is increased to 0.3 eV, the opposite direction for having a band inversion at $\Gamma$. With a FM interlayer coupling (see Fig.\,\ref{fig:Band}d), the band double degeneracy is lifted and the top valence band and bottom conduction band move toward each other reducing the band gap to 0.2 eV, albeit still no band inversion. Furthermore, from the total density of states (see Fig.~S9 in Supporting Information) for all the site-mixing supercell configurations considered with both AFM and FM-like coupling, there is a bulk band gap of 0.2 eV or beyond. Compared to the FM configuration of ideal  MnSb$_2$Te$_4$, in contrast to the increased band gap at $\Gamma$ and Z points, the gap at L point in Fig.\,\ref{fig:Band}(d) is reduced for MnSb$_2$Te$_4$ with significant site mixing. These band structure results indicate that although site-mixing in MnSb$_2$Te$_4$ is beneficial to stabilize a FM interlayer coupling, it is detrimental to retain the band inversion needed for non-trivial topology. A high magnetic field is expected to ferromagnetically align Mn moments at both Mn and Sb sites. As shown in Fig.~\ref{fig:DFT2} in Supporting Information, the calculation with all moments ferromagnetically aligned suggests still no band inversion although the band gap is slightly reduced compared to that in Fig.\,\ref{fig:Band}(d). To justify the calculations, Fig.~\ref{fig:STS} in Supporting Information shows the scanning tunneling spectra data of T$_C$24K and T$_N$19K at 4.5\,K. Both samples have a bulk band gap of about 0.4~eV. The flat and diminished density of state within the bandgap indicates the absence of in-gap states and trivial topology of the band structures. The calculated band gap in these supercell configurations agrees with the scanning tunneling spectra (STS) data, considering the underestimation of band gap of the PBEsol exchange-correlation functional.

\section{Discussion}
\subsection{Effects of site mixing on the magnetic and electronic properties of MnSb$_2$Te$_4$}

Our results show that magnetic defects can have a dramatic effect on the magnetic ordering in MnSb$_2$Te$_4$, and Mn-Sb site mixing favors a FM interlayer coupling. The anisotropic magnetic properties shown in Fig.\,\ref{fig:Mag} demonstrate that the inter-septuple-layer coupling can be FM or AFM depending on the crystal growth parameters. Our neutron diffraction and microscopy measurements found a significant amount of Mn-Sb site mixing with nearly random distribution in all crystals investigated in this work. We noticed that similar Mn-Sb site mixing is also observed in a polycrystalline sample with a FM interlayer coupling by neutron powder diffraction~\cite{murakami2019realization} and in a single crystal sample by single-crystal x-ray diffraction~\cite{zhou2020topological}. The observation of Mn-Sb site mixing in MnSb$_2$Te$_4$ samples synthesized differently by different groups suggest that the presence of Mn-Sb mixing is a general phenomena in MnSb$_2$Te$_4$, possibly due to the small formation energy of antisite defects~\cite{du2020tuning}. By investigating the anisotropic magnetic properties, concentration and distribution of Mn-Sb site mixing, this work develops a direct correlation between the magnetic ground states and magnetic defects. Other intrinsic lattice defects, such as Sb$^{,}_{Te}$ and Te$^{.}_{Sb}$, might also exist with a smaller concentration and mainly affect the transport properties. Therefore, the formation of the complex magnetic structure and the sign change of inter-septuple-layer coupling result from the occupation of Mn at Sb sites. We noticed that with increasing growth temperature, both Mn$_{Mn}^\times$ and Mn$_{Sb}'$ increase in concentration, and therefore does the total Mn content. The difference between T$_C$24K and T$_N$19K suggests a threshold concentration of $~$13\% Mn$_{Sb}'$, above which the inter-septuple-layer interaction becomes FM. With further increasing the total Mn content, the magnetic ordering temperature increases to about 50\,K as illustrated in Fig.~\ref{fig:Tunable} in Supporting Information. Therefore, both the sign of interlayer coupling and the magnetic ordering temperature are tunable by a careful control of the concentration and distribution of magnetic ions in the lattice.

The modification of the magnetic ground state results from the extra competing interactions introduced by Mn$_{Sb}'$. The energy difference between FM and AFM states is small in both ideal MnBi$_2$Te$_4$ and MnSb$_2$Te$_4$. Therefore, the magnetic ground state can be sensitive to external perturbations.  The competing interactions in MnBi$_2$Te$_4$ were studied previously using inelastic neutron scattering~\cite{li2020competing}. Similar competing interactions are expected in ideal MnSb$_2$Te$_4$. The presence of Mn$_{Sb}'$ introduces extra competing magnetic interactions. Considering a FM order is always observed in \textit{TM}-doped (\textit{TM}=Mn, Cr, and V) Bi$_2$Te$_3$ and Sb$_2$Te$_3$, the interaction between neighbouring Mn$_{Sb}'$ and Mn$_{Sb}'$ ions right across the van der Waals gap is expected to be FM. This FM interaction increases in magnitude with increasing amount of Mn$_{Sb}'$ and competes with the AFM interaction in between Mn$_{Mn}^\times$ and Mn$_{Mn}^\times$ in neighbouring septuple layers and eventually leads to the sign change of interlayer magnetic coupling. Other interactions between further neighbouring Mn ions will also contribute to the overall interlayer magnetic interactions but are possibly less important. Mn$_{Sb}'$ magnetic defects also introduce extra exchange interactions in each septuple layer. Both 90 degree and 180 degree Mn-Te-Mn superexchange interactions are active and contribute to the overall interaction between Mn$_{Sb}'$ and Mn$_{Mn}^\times$. While the partial contribution of different exchange paths to the total exchange are to be investigated, the experimentally observed ferrimagnetic arrangement in each septuple layer suggests an AFM interaction between Mn$_{Sb}'$ and Mn$_{Mn}^\times$. As described later, this AFM interaction is rather strong.

Mn-Sb site mixing favors a FM interlayer coupling but is detrimental to the band inversion. Our DFT calculations considered different antisite configurations using the (2$\times$2$\times$2) supercell and found all antisite configurations prefer a FM interlayer coupling. This is further confirmed by using a larger supercell (3$\times$3$\times$2). This FM interlayer coupling is preferred for the realization of a Weyl state. Unfortunately, our band structure calculations suggest a sizable band gap of $>$0.2\,eV regardless of the sign of interlayer magnetic interaction or the antisite configuration. STS measurement further confirms the sizable gap in our crystals with significant amount of Mn-Sb site mixing. The site-mixing induced gap opening is also reproduced in a most recent study by Wimmer et al~\cite{wimmer2020ferromagnetic}. In their DFT calculations, they considered the effect of 20\% Mn-Sb site mixing  on the band structure and found a gap of about 0.2\,eV which makes MnSb$_2$Te$_4$ to be topological trivial. In addition, Wimmer et al. also considered the effect of 5\% Mn-Sb antisite mixing on the band structure of FM MnSb$_2$Te$_4$ to find that the band inversion and Weyl point are preserved. This interesting result indicates that it is likely to tune the magnetism by introducing Mn-Sb (or Mn-Bi) site mixing in MnSb$_2$Te$_4$ (or MnBi$_2$Te$_4$) without losing the band inversion as suggested below.

\subsection{Indications for MnTe(Bi$_2$Te$_3$)$_n$}

Since about 3\% Mn$_{Bi}'$ ions were observed in MnBi$_2$Te$_4$ single crystals by different groups, one naturally asks (1) whether those 3\% Mn$_{Bi}'$ ions order magnetically forming ferrimagnetic septuple layers as observed in MnSb$_2$Te$_4$, (2) whether it is possible to tune the magnetic ground state, and thus the topological properties, by controlling the concentration and distribution of magnetic defects in MnBi$_2$Te$_4$, and (3) whether the site mixing plays any role in other n$>$1 compounds in MnTe(Bi$_2$Te$_3$)$_n$ family with more complex stacking of the septuple and quintuple layers. As discussed below, Mn$_{Bi}'$ ions likely order magnetically as observed in MnSb$_2$Te$_4$ and the magnetism of Mn$_{Bi}'$, in either the septuple or the quintuple layers, should be considered in all MnTe(Bi$_2$Te$_3$)$_n$ compounds.

We noticed that the growth dependent magnetic behavior as shown in Fig.\,\ref{fig:Mag} is absent in MnBi$_2$Te$_4$. All previous work reported an AFM order with T$_N$=25\,K for MnBi$_2$Te$_4$ crystals. This consistency might result from the narrow crystallization temperature window and/or the limited concentration of magnetic defects due to the large difference of ionic size and electronegativity between Mn and Bi. Considering possible AFM alignment of Mn$_{Bi}'$  and Mn$_{Mn}^\times$, the field dependence of magnetization can provide more valuable information.

As proposed by Murakami et al.~\cite{murakami2019realization}, the ferrimagnetic arrangement of Mn moments in each septuple layer of MnSb$_2$Te$_4$ provides a straightforward explanation for the large difference between the ordered and saturation moments. With the ordered moment at 6.5\,K and the occupancy both determined from neutron single crystal diffraction, the average moment within each septuple layer is 1.52(4) and 1.65(8)$\mu_{B}$/Mn at 6.5\,K for T$_N$19K and T$_C$24K, respectively (Table\,S6 in Supporting Information). These values are smaller than the saturation moment determined at 2\,K from magnetic measurements but in reasonable agreement. The magnetization at 2\,K does not saturate even in an applied magnetic field of 120\,kOe, indicating a strong magnetic coupling between Mn$_{Mn}^\times$ and Mn$_{Sb}'$. A field-induced transition to a state with all Mn moments polarized is expected at a much higher magnetic field and this deserves further study.

One important indication of the above analyses is possible magnetic order of Mn$_{Bi}'$ ions in MnBi$_2$Te$_4$. The saturation moment of MnBi$_2$Te$_4$ is $\approx$3.8\,$\mu_{B}$/Mn estimated from magnetic measurements at 2\,K~\cite{yan2019crystal,yan2019evolution,li2020competing}, which is smaller than the ordered moment of 4.04(13)\,$\mu_{B}$/Mn at 10\,K~\cite{yan2019crystal} or 4.7(1)\,$\mu_{B}$/Mn at 4.5\,K~\cite{ding2020crystal} determined from neutron diffraction. The amount of Mn$_{Bi}'$ is $\approx$3\% in our crystals~\cite{yan2019crystal}. With the assumption that Mn$_{Mn}^\times$ and Mn$_{Bi}'$ ions are antiferromagnetically aligned with the same ordered moment of 4.7\,$\mu_{B}$, the net moment of each septuple layer is about 4.1\,$\mu_{B}$, in reasonable agreement with the experimentally measured saturation moment at 2\,K. It should be noted that 2\% Mn on Bi sites is enough to induce a long range FM order in Bi$_2$Te$_3$. From these points of view, the magnetic structure of MnBi$_2$Te$_4$ and its field dependence should be carefully revisited. Considering the similarity between MnBi$_2$Te$_4$ and MnSb$_2$Te$_4$, a similar complex ferrimagnetic structure is expected for MnBi$_2$Te$_4$ if Mn$_{Bi}'$ ions forms a long range magnetic order below T$_N$=25\,K. The doping dependence of magnetization in MnBi$_{2-x}$Sb$_x$Te$_4$ was previously carefully studied~\cite{yan2019evolution}. The gradually suppressed saturation moment with increasing Sb doping signals the increasing amount of Mn at Bi/Sb site. The evolution with doping of the magnetic and topological properties should be revisited in this system by considering the doping-dependent site mixing and the magnetic ordering of antisite Mn ions. The magnetic order of antisite Mn ions will impact the surface magnetism and, of particular interest are whether and how this magnetic order couples to the topological surface states which deserve further investigation. Similar studies are also relevant for MnTe(Bi$_2$Te$_3$)$_n$ compounds with n$>$1 considering the presence of Mn$_{Bi}'$ in both the quintuple and septuple layers. STM studies observed 2-3\% Mn$_{Bi}'$ ions in both the quintuple and septuple layers in MnBi$_4$Te$_7$ crystals~\cite{liang2020mapping,wu2020distinct}.

Besides the natural antisite defects, one can purposely introduce magnetic ions, such as V, or Cr, at Bi site to tune the magnetism and even changing the sign of inter-septuple-layer magnetic interaction in MnBi$_2$Te$_4$. Realization of such a complex magnetic structure without losing the band inversion might be a valid approach toward QAHE in MnBi$_2$Te$_4$. Considering the narrow crystallization temperature window for the flux growth of MnBi$_2$Te$_4$ out of Bi-Te flux, vapor transport growth of bulk crystals or MBE growth of thin films might have a better control of the concentration of magnetic ions on Bi sites. On the other hand, reducing the amount of site mixing might lead to quantized transport properties at even higher temperatures than reported in MnBi$_2$Te$_4$ flakes~\cite{deng2020quantum,liu2020robust,ovchinnikov2020intertwined}.

\section{Summary}
In summary, we have demonstrated the importance of understanding and controlling  the defect concentration and distribution in any candidate magnetic topological insulators. In particular, we have shown that Mn-Sb site-mixing can have a dramatic effect on the magnetic ground state and electronic structure of MnSb$_2$Te$_4$. These antisite defects favor a FM interlayer coupling. While an ideal FM MnSb$_2$Te$_4$ crystal hosts Weyl points, site-mixing is detrimental to the desired nontrivial band topology. Our work highlights the importance of further investigating lattice defects in the isostructural MnBi$_2$Te$_4$ and related compounds, where Mn-Bi site mixing has been observed although by a much reduced amount. How the magnetic defects Mn$_{Bi}'$ in MnTe.nBi$_2$Te$_{3}$ affect the surface magnetism and interact with the topological surface states deserve further study.

\section{Acknowledgment}
The authors thank Huibo Cao, Maohua Du, Robert McQueeney, Satoshi Okamoto, Xiaodong Xu for helpful discussions and Michael Lance for WDS measurements. Research at ORNL (MM, BS, and JY) was supported by the US Department of Energy, Office of Science, Basic Energy Sciences, Division of Materials Science and Engineering. LW was supported by the Center for the Advancement of Topological Semimetals, an Energy Frontier Research Center funded by U.S. Department of Energy (DOE), Office of Science, Basic Energy Sciences. Ames Laboratory is operated for the US Department of Energy by Iowa State University under Contract No. DE-AC02-07CH11358. QZ acknowledges support from the Gordon and Betty Moore Foundation’s EPiQS Initiative, Grant GBMF9069. ZH and WW were supported by NSF Grant No. DMR-1506618 and EFMA-1542798. MC is supported by an Early Career project supported by DOE Office of Science. A portion of this research used resources at the High Flux Isotope Reactor, Spallation Neutron Source and the Center for Nanophase Materials Sciences, DOE Office of Science User Facilities operated by the Oak Ridge National Laboratory.

\section{references}

\newpage

\newcolumntype{C}[1]{>{\centering\let\newline\\\arraybackslash\hspace{0pt}}m{#1}}
\section{Supporting Information} 
\renewcommand\thefigure{S\arabic{figure}}
\renewcommand\thetable{S\arabic{table}}
\renewcommand\thesection{S\arabic{section}}

\setcounter{figure}{0}

\subsection{Single Crystal Neutron Diffraction}
Single crystal neutron diffraction data were collected at 300~K using the beamline TOPAZ at the Spallation Neutron Source (SNS) to determine the average nuclear structure. Low temperature measurements were performed at the beamline CORELLI at SNS and the Four-Circle Diffractometer (HB-3A) at the High Flux Isotope Reactor (HFIR) to determine the average magnetic structure.

\begin{figure*}[tb] \centering \includegraphics [width = \textwidth] {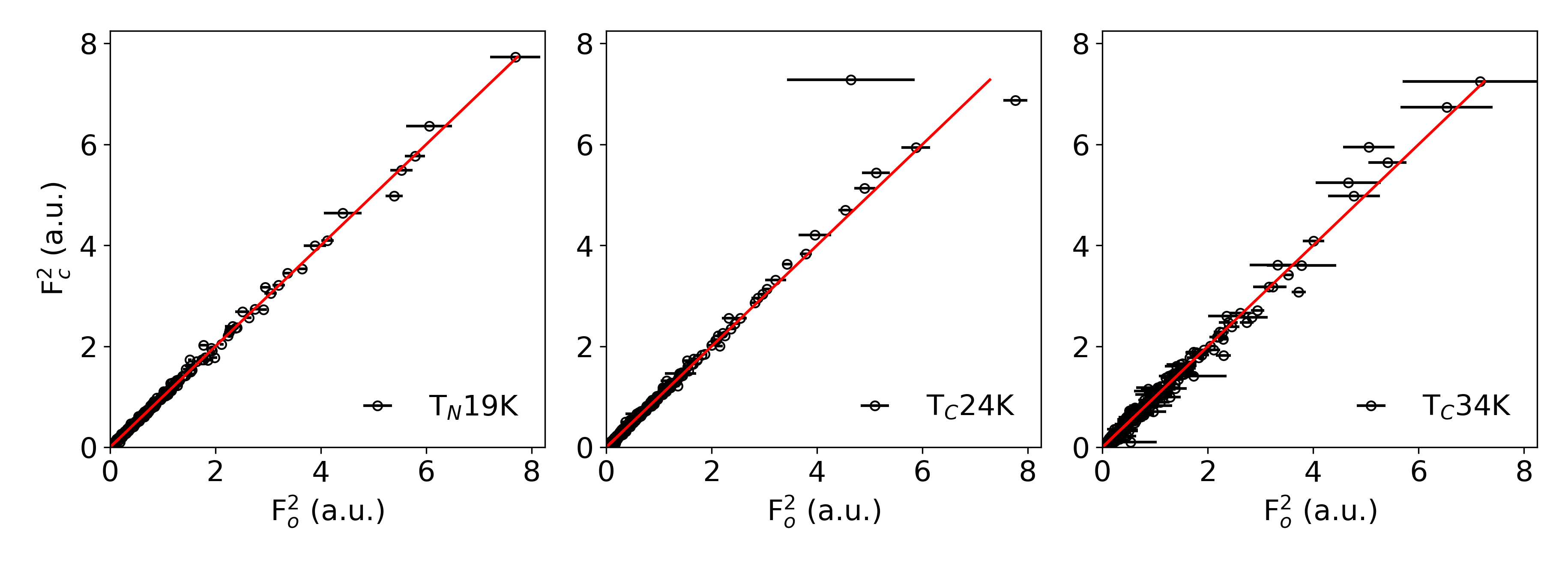}
\caption{(color online) The observed squared structure factors vs the calculated ones from the nuclear structure refinements using the space groups $R\bar{3}m$ for (a) T$_{N}$19K, (b) T$_{C}$24K and (c) T$_{C}$34K, respectively. R1=3.22\%, 5.33\% and 5.76\% for T$_{N}$19K, T$_{C}$24K and T$_{C}$34K, respectively.}
\label{fig:Rn3m}
\end{figure*}

\subsubsection{Nuclear structure Refinement}
Previous studies reported the $R\bar{3}m$ space group for MnSb$_2$Te$_4$ from powder diffraction experiments~\cite{yan2019evolution, murakami2019realization}. With the high resolution single crystal neutron diffraction data collected at TOPAZ (300\,K), we examined the crystal structures of all three MnSb$_2$Te$_4$ single crystals. Figure~\ref{fig:Rn3m} shows the refinement results using $R\bar{3}m$ for $T_{N}19$K, $T_{C}24$K and T$_{C}$34K. Lowering the crystal symmetry from $R\bar{3}m$ to its subgroup \textit{R}3\textit{m} or \textit{C}2/\textit{m} does not significantly improve the refinement; rather there are high correlated atomic positions and atomic displacement parameters, suggesting that a lower symmetry is unnecessary to describe the average structure in the presence of significant Mn-Sb site mixing. The $R\bar{3}m$ space group is thus used for all structure refinements below.

\begingroup
\begin{table*}[tb]
%\small
\begin{tabular}{|C{2.8cm}|C{1.3cm}|C{1.3cm}|C{1.38cm}||C{1.38cm}|C{1.3cm}|C{1.3cm}||C{1.3cm}|C{1.3cm}|C{1.3cm}|}
\hline
                 & \multicolumn{3}{c||}{T$_{N}$19K} & \multicolumn{3}{c||}{T$_{C}$24K} & \multicolumn{3}{c|}{T$_{C}$34K}                                              \\
\hline
total peaks ($I \geq 2 \delta_{I}$)             & \multicolumn{3}{c||}{7086} & \multicolumn{3}{c||}{6151}& \multicolumn{3}{c|}{1976}              \\
\hline
unique peaks for R1 ($I \geq 4 \delta_{I})$ & \multicolumn{3}{c||}{705} & \multicolumn{3}{c||}{666}& \multicolumn{3}{c|}{529}              \\
\hline
Assumed Formula & A & B & C & A & B & C & A & B & C\\
\hline
\# of parameters (constraints) & 15(0) & 15(1) & 16(1) &  15(0) & 15(1) & 16(1)& 15(0) & 15(1) & 16(1) \\
\hline
R1   & 3.22\% & 3.22\% & 3.22\% &  5.33\% & 5.34\% & 5.34\% & 5.76\% & 5.76\% & 5.76\% \\
\hline
Sb$_{Sb}$  & 0.871(2) & 0.871(2) & 0.871(2) & 0.849 (2) & 0.847(2) & 0.847(2) & 0.842(3) & 0.842(3) & 0.842(3) \\
\hline
Mn$_{Sb}'$  & 0.129(2) & 0.129(2) & 0.129(2) & 0.151(2) & 0.153(2) & 0.153(2) & 0.158(3) & 0.158(3) & 0.158(3) \\
\hline
Mn$_{Mn}$  & 0.588(2) & 0.588(2) & 0.549 (4) & 0.645(2) & 0.648(4) & 0.637(6) & 0.674(6) & 0.674(4) & 0.672(6) \\
\hline
Sb$_{Mn}^.$  & 0.412(2) & 0.412(2) & 0.387 (2) & 0.335 (2) & 0.352(4) & 0.344(4) & 0.326(3) & 0.326(4) & 0.324(3) \\
\hline
Te$_{Te}$ & 1 & 0.962(3) & 1 & 1 & 0.989(3)& 1 &1 & 0.997(3) & 1\\
\hline
Sb$_{Te}'$ & 0 & 0.038(3) & 0 & 0 & 0.011(3)& 0 &0 & 0.003(3) & 0\\
\hline

\end{tabular}
\caption{Comparison of the refinement quality for all three crystals using different assumed chemical formulas. Formula A: (Mn$_{1-x}$Sb$_x$)(Sb$_{1-y}$Mn$_y$)$_2$Te$_4$, formula B: (Mn$_{1-x}$Sb$_x$)(Sb$_{1-y}$Mn$_y$)$_2$(Te$_z$Sb$_{1-z}$)$_4$, and formula C: (Mn$_x$Sb$_y$)(Sb$_{1-z}$Mn$_z$)$_2$Te$_4$. Charge neutrality was assumed for formulas B and C using the nominal valance state, Mn$^{2+}$ at Mn and Sb sites, Sb$^{3+}$ at Mn and Sb sites, Sb$^{3-}$ at Te sites, and Te$^{2-}$ at Te sites.}
\label{tab:StruRefine}
\end{table*}
\endgroup

\subsubsection{Mn-Sb site mixing}
Table~\ref{tab:StruRefine} shows the comparison of the structural refinements on the 300~K TOPAZ data assuming different chemical formula. All have the same fitting quality.  However,  the refinements suggest that all three samples have significant amount of Mn-Sb site mixing with slight differences in the numerical values. Importantly, they all found more Mn ions on both Mn and Sb sites for samples with a ferromagnetic interlayer coupling. Detailed structural refinement results are shown in Tables~\ref{tab:ND_AFM19K},~\ref{tab:ND_FM24K},~\ref{tab:ND_FM34K} for T$_{N}$19K, T$_{C}$24K and T$_{C}$34K,   respectively. With increasing Mn content, \textit{a}-lattice parameter decreases slightly while \textit{c}-lattice is less affected.

\begin{table}[tb]
\centering
\begin{tabular}{c|p{0.6cm}p{0.6cm}p{1.8cm}|c|c|c}
  \hline
  \hline
  Atom   &      $x$  &  $y$ &    $z$       &  occ.     & U$_{eq}$ ($\AA^2$)    &  Wyck.\\
 \hline
  Te1    &       0   & 0  &  0.13162(2)    & 1         &  0.01710(8)   & $6c$  \\
  Te2    &       0   & 0  &  0.29212(2)    & 1         &  0.02922(2)   & $6c$  \\
  Sb     &       0   & 0  &  0.42559(2)    & 0.871(2)  &  0.02028(13)  & $6c$   \\
  Mn$_{Sb}'$  & 0   & 0  &  0.42559(2)    & 0.129(2)  &  0.02028(13)  & $6c$  \\
  Mn     &       0   & 0  &  0             & 0.588(4)  &  0.008(5)     & $3a$  \\
  Sb$_{Mn}^.$  & 0   & 0  &  0             & 0.412(2)  &  0.008(5)     & $3a$  \\

 \hline
 \end{tabular}
\caption{Refined structural parameters of T$_{N}$19K. The composition determined from the best fit is Mn$_{0.84}$Sb$_{2.16}$Te$_4$ and the lattice parameters are $a~=~4.2543(1)$~\AA, $c=40.899(2)$~\AA, and V~=~641.07(4)~\AA$^3$.}
\label{tab:ND_AFM19K}
\end{table}

\begin{table}[tb]
\centering
\begin{tabular}{c|p{0.6cm}p{0.6cm}p{1.8cm}|c|c|c}
  \hline
  \hline
  Atom   &        $x$     &       $y$     &      $z$       & occ.&  U$_{eq}$ ($\AA^2$)   &  Wyck.\\
 \hline
  Te1   &       0   & 0  &  0.13140(2)   & 1     &  0.01780(12)  & $6c$  \\
  Te2   &       0   & 0  &  0.29230(2)   & 1     &  0.01725(12)    & $6c$   \\
  Sb    &       0   & 0  &  0.42567(2)   & 0.850(2)    & 0.01888(19)    & $6c$   \\
  Mn$_{Sb}'$  &        0   & 0  &  0.42567(2)   & 0.150(2)     &  0.01888(19)    & $6c$   \\
  Mn    &       0   & 0  &  0            & 0.635(2)  & 0.01888(19)    & $3a$  \\
  Sb$_{Mn}^.$  &        0   & 0  &  0            & 0.365(2)  &  0.01888(19)    & $3a$  \\
 \hline
 \end{tabular}
\caption{Refined structural parameters of  T$_{C}$24K. The composition determined from the best fit is Mn$_{0.93}$Sb$_{2.07}$Te$_4$. The lattice parameters are $a~=~4.2510(2)$~\AA~, $c=40.907(3)$~\AA, and V~=~640.17(6)~\AA$^3$ .}
\label{tab:ND_FM24K}
\end{table}

\begin{table}[tb]
\centering
\begin{tabular}{c|p{0.6cm}p{0.6cm}p{1.8cm}|c|c|c}
  \hline
  \hline
  Atom   &        $x$     &       $y$     &      $z$       & occ.&  U$_{eq}$ ($\AA^2$)   &  Wyck.\\
 \hline
  Te1   &       0   & 0  &  0.13127(2)   & 1                &  0.01518(18)  & $6c$  \\
  Te2   &       0   & 0  &  0.29244(2)   & 1                &  0.01455(18)    & $6c$   \\
  Sb    &       0   & 0  &  0.42571(3)   & 0.842(3)    & 0.0156(3)    & $6c$   \\
  Mn$_{Sb}'$  &        0   & 0  &  0.42571(3)   & 0.158(3)    &  0.0156(3)    & $6c$   \\
  Mn    &       0   & 0  &  0                 & 0.674(4)    & 0.017(2)    & $3a$  \\
  Sb$_{Mn}^.$  &        0   & 0  &  0                  & 0.326(4)    &  0.017(2)    & $3a$  \\
 \hline
 \end{tabular}
\caption{Refined structural parameters of  T$_{C}$34K. The composition determined from the best fit is Mn$_{0.99}$Sb$_{2.01}$Te$_4$ $a~=~4.2406(7)$~\AA~, $c=40.826(10)$~\AA~and V~=~635.8(3)~\AA$^3$ .}
\label{tab:ND_FM34K}
\end{table}

\subsubsection{neutron diffraction patterns of T$_C$24K above Tc}

\begin{figure*} \centering \includegraphics [width = 0.72\textwidth] {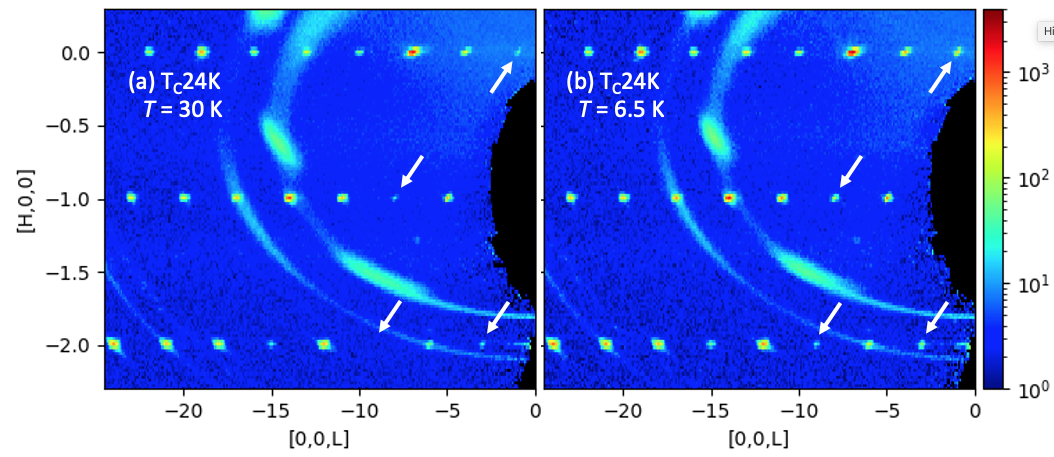}
\caption{(color online) Neutron scattering pattern in the [H~1~L] plane of the T$_C$24K sample at (a) 30\,K and (b) 6.5\,K. The white arrows highlight those Bragg peaks that have small nuclear scattering contributions and gain significant intensities from magnetic diffraction below $T_{C}$. The ring- and arc-like features are from the sample environment and sample mount.}
\label{fig:CORT30Kvs6K}
\end{figure*}

Figure\,\ref{fig:CORT30Kvs6K} shows the neutron scattering patterns in the [H~1~L] plane of the T$_C$24K sample collected at (a) 30\,K and (b) 6.5\,K. The observed Bragg peaks match the scattering condition -H+K+L = 3n, as required by the space group $R\bar{3}m$ for the nuclear reflections. The pattern collected at 6.5\,K is also presented in the main text and it is presented here to highlight the change cooling below $T_C$. The diffraction pattern collected at 6.5\,K is similar to that collected at 30\,K but the intensity of some peaks is enhanced as highlighted by the white arrows, suggesting a magnetic wavevector of $k_{FM}~=~(0~0~0)$.

\subsubsection{magnetic structures: ordered moments on Mn$_{Sb}'$}
\begin{figure} \centering \includegraphics [width = 0.48\textwidth] {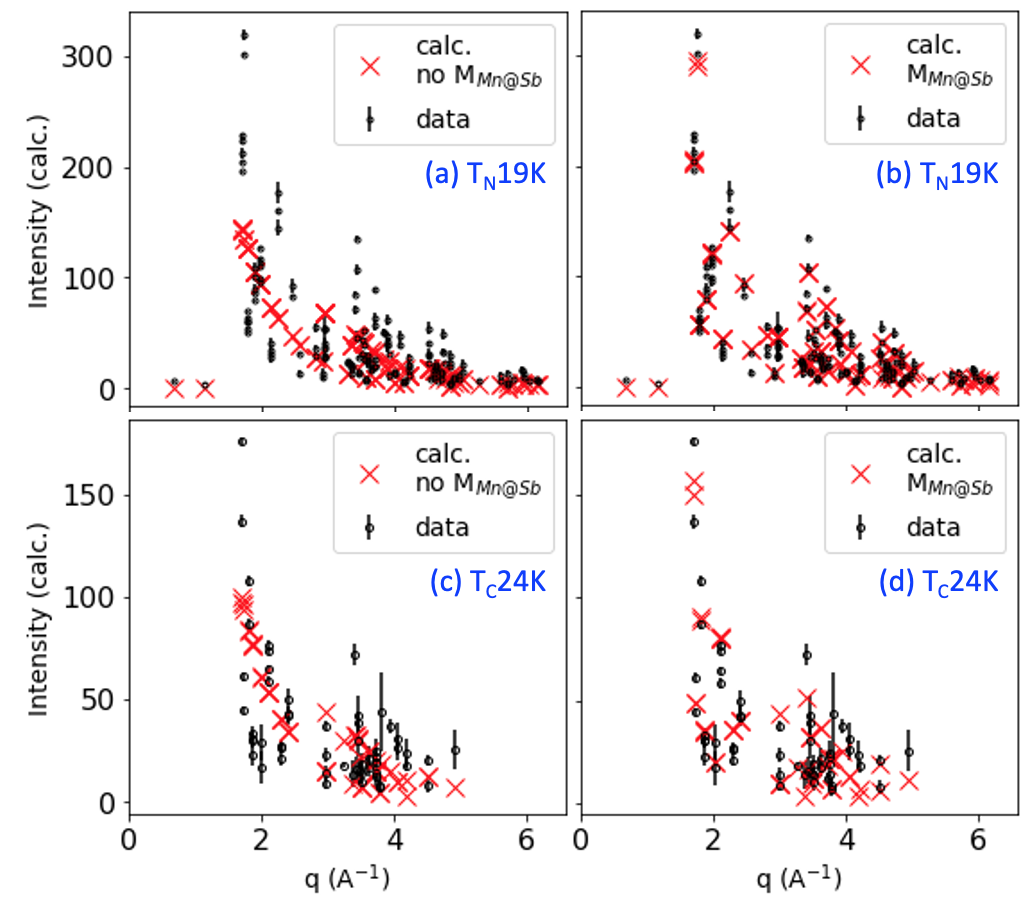}
\caption{(color online) Comparison between the magnetic structure model without (a),(c) and with (b),(d) considering ordered moments on Mn$_{Sb}$ sites for the T$_{N}$19K and T$_{C}$24K. Refinement results suggests ordered moments of Mn$_{Sb}$ sites,  which are antiparallel to the ordered moments on adjacent Mn$_{Mn}$sites within each septuple layer.}
\label{fig:magstru}
\end{figure}

Reciprocal space mapping from the CORELLI data found no highly structured diffuse scattering signal, which suggests that MnSb$_2$Te$_4$ remains in the $R\bar{3}m$ space group at low temperatures. Low-temperature datasets were collected for T$_{N}$19K and T$_{C}$24K. Partially due to the unavoidable background scattering from the sample environment, the datasets collected at both CORELLI and HB-3A contain much fewer peaks than the 300\,K TOPAZ datasets. However, the average nuclear structures determined from the CORELLI and HB-3A data for T$_{N}$19K and T$_{C}$24K suggest similar amounts of Mn/Site site mixing but with higher uncertainties in comparison to the TOPAZ results. Table~\ref{tab:CompND_AFM19K} shows the site mixing and ordered moments for T$_N$19K refined using the formula A from neutron diffraction data collected at three different beamlines. Interestingly, to achieve satisfactory refinements of the magnetic structure, one has to consider the ordered moments on Mn$_{Sb}'$, as shown in Fig.~\ref{fig:magstru} for both T$_{N}$19K and T$_{C}$24K. For T$_{N}$19K, the ordered moments are 3.71(2)~$\mu_B$ per Mn$_{Mn}$ and 3.46(8)~$\mu_B$ per Mn$_{Sb}'$ using the chemical formula A and 3.95(2)~$\mu_B$ per Mn$_{Mn}$ and 3.18(7)~$\mu_B$ per Mn$_{Sb}'$ using the chemical formula C. For T$_{C}$24K, the ordered moments are 4.08(7)~$\mu_B$ per Mn$_{Mn}$ and 3.5(2)~$\mu_B$ per Mn$_{Sb}$ for the chemical formula A and 4.12(7)~$\mu_B$ per Mn$_{Mn}$ and 3.3(2)~$\mu_B$ per Mn$_{Sb}'$ for the chemical formula C. The determined ordered moments using different chemical formulas are slightly different but agree with each other. We thus use formula A in the main text to describe the chemical composition and occupancy. Table\,\ref{tab:TOPAZ} summarizes the nonstoichiometry, site occupancy, and magnetic moments for all three crystals.

\begin{table}[tb]
\centering
\begin{tabular}{|C{2.4cm}|C{1.8cm}|C{1.8cm}|C{1.8cm}|}
\hline
     & TOPAZ (300~K) &   CORELLI (6.5~K)    & Four-Circle (4.5~K)\\
 \hline
 nuclear  peaks ($I~\geq~2\delta_{I}$)    & 7086 & 189 & 96 \\
 \hline
 magnetic  peaks ($I~\geq~2\delta_{I}$)    & 0 & 153 & 35 \\
 \hline
 Sb$_{Sb}$ & 0.871(2) & 0.881(6) & 0.88(1) \\
 \hline
 Mn$_{Sb}'$, $p_2$ & 0.129(2) & 0.119(6) & 0.12(1) \\
 \hline
 Mn$_{Mn}$, $p_1$  & 0.588(2) & 0.587(7) & 0.58(1) \\
\hline
Sb$_{Mn}^.$& 0.412(2) & 0.414(7) & 0.42(1) \\
\hline
Mn$_{Mn}$ ($\mu_B$/Mn) & - & 3.71(2) & 3.9(1) \\
\hline
Mn$_{Sb}'$ ($\mu_B$/Mn) & - & 3.46(8) & 2.7(3) \\
\hline
\end{tabular}
\caption{Comparison on the refined values on the antisite mixing and ordered moments on T$_{N}$19K from the three neutron diffractometers using the formula A: (Mn$_{p1}$Sb$_{1-p1}$)(Sb$_{1-p2}$Mn$_{p2}$)$_2$Te$_4$. }
\label{tab:CompND_AFM19K}
\end{table}

\begin{table*}[htbp]
\caption{Nonstoichiometry, site occupancy, and magnetic moments for all three crystals. The chemical formula is written as (Mn$_{p1}$Sb$_{1-p1}$)(Sb$_{1-p2}$Mn$_{p2}$)$_2$Te$_4$, where Te occupancy was fixed to be 1 and $p_1$ and  $p_2$ are the site occupation of Mn atoms at Mn and Sb sites, respectively, determined from the 300\,K TOPAZ data.  The ordered moments (in $\mu_B$/Mn) are determined from the 6.5\,K CORELLI data. The average moment within each septuple layer in the ferrimagnetic state was estimated as $ (m_{Mn@Mn} \times p_{1}- m_{Mn@Sb} \times 2p_2)/(p_{1} + 2p_2)$. }
\centering
\begin{tabular}{c|c|c|c}
  \hline
  \hline
  & T$_N$19K &  T$_C$24K & T$_C$34K \\
  \hline
 WDS/EDX & Mn$_{0.82}$Sb$_{2.17}$ (WDS)  &  Mn$_{0.90}$Sb$_{2.10}$ (WDS) & Mn$_{0.99}$Sb$_{2.01}$ (EDS) \\
  \hline
 neutron     & Mn$_{0.84}$Sb$_{2.16}$  &  Mn$_{0.93}$Sb$_{2.07}$  &  Mn$_{0.99}$Sb$_{2.01}$  \\
  \hline
  $p_1$, Mn$_{Mn}^\times$  & 0.588(2)& 0.635(2) & 0.674(4) \\
  \hline
  $p_2$, Mn$_{Sb}'$ & 0.129(2)   & 0.150(2)  & 0.158(3) \\
  \hline
  Ordered moment at 6.5K, Mn$_{Mn}^\times$  & 3.71(2) & 4.08(7)& - \\
  \hline
  Ordered moment at 6.5K, Mn$_{Sb}'$& 3.46(8) & 3.5(2) & -  \\
  \hline
Estimated M$_{septuple}$ & 1.52(4)   & 1.65(8)  & - \\
  \hline
 Saturation moment at 2\,K, Ms  & 1.95(2)   & 2.00(2)  & 1.98(2) \\
  \hline
 \end{tabular}
\label{tab:TOPAZ}
\end{table*}

\begin{figure*} \centering \includegraphics [width = 0.8\textwidth] {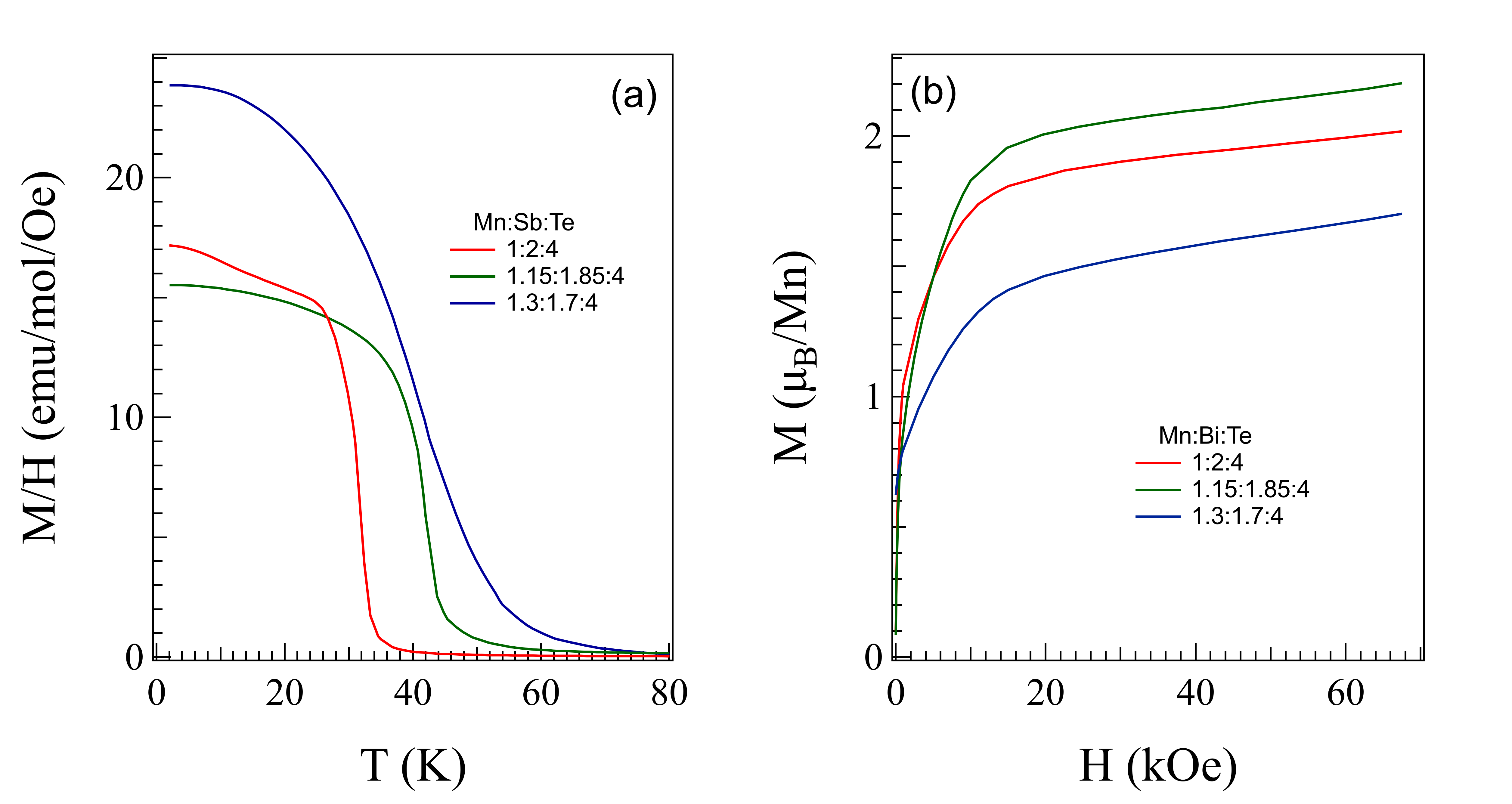}
\caption{(color online) (a) Temperature dependence of magnetic susceptibility of crystals measured in an applied magnetic field of 100 \,Oe in a field-cooled mode. (b) Field dependence of magnetization at 2\,K.}
\label{fig:Tunable}
\end{figure*}

\subsection{Tunable magnetic ordering temperatures}

Figure\,\ref{fig:Tunable}a shows the temperature dependence of magnetic susceptibility of MnSb$_2$Te$_4$ synthesized using different ratios of starting elements.  All three batches are homogenized at 1000~$^\circ$C for 12 hours and then annealed at 640~$^\circ$C for a week. The atomic ratio of the starting composition is detailed in the figure. Changing the Mn-Sb ratio is expected to vary the magnetic properties. MnSb$_2$Te$_4$ crystals grown in this manner are rather small and the magnetic measurements were performed using a cluster of small crystals with a random orientation. The results suggest that $T_C$ can be tuned up to around 50\,K. We noticed that there is a small fraction of unknown magnetic impurity for the composition starting with Mn:Sb=1.3:1.7. Further increasing Mn content in the starting materials increases the content of the impurity phase.

Figure \,\ref{fig:Tunable}b shows the magnetization measured at 2K in applied magnetic fields up to 70\,kOe. It is interesting that all samples have a magnetization around 2~$\mu_B$/Mn at 2\,K and 70\,kOe. While detailed nonstoichiometry and atomic occupancy need to be determined, it is reasonable to expect that they determine the ordering temperature and saturation moment. The sample with $T_C$ around 50\,K has a smaller magnetization possibly due to the presence of the unknown magnetic impurity.

\begin{figure} \centering \includegraphics [width = 0.46\textwidth] {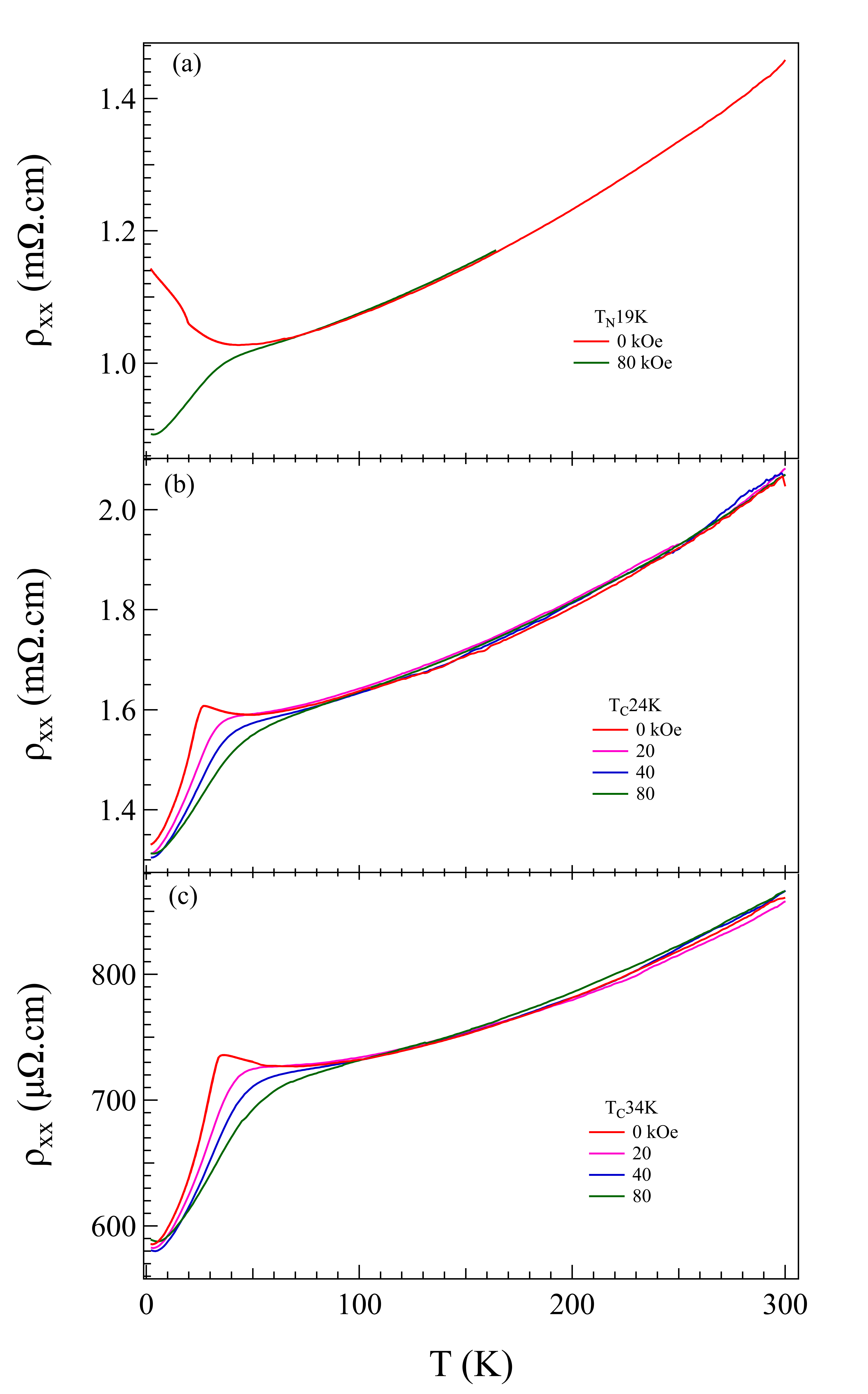}
\caption{(color online) Temperature dependence of in-plane electrical resistivity with the electrical current flowing in the \textit{ab}-plane and a magnetic field applied along the crystallographic \textit{c}-axis. }
\label{fig:Rxx}
\end{figure}

\begin{figure} \centering \includegraphics [width = 0.46\textwidth] {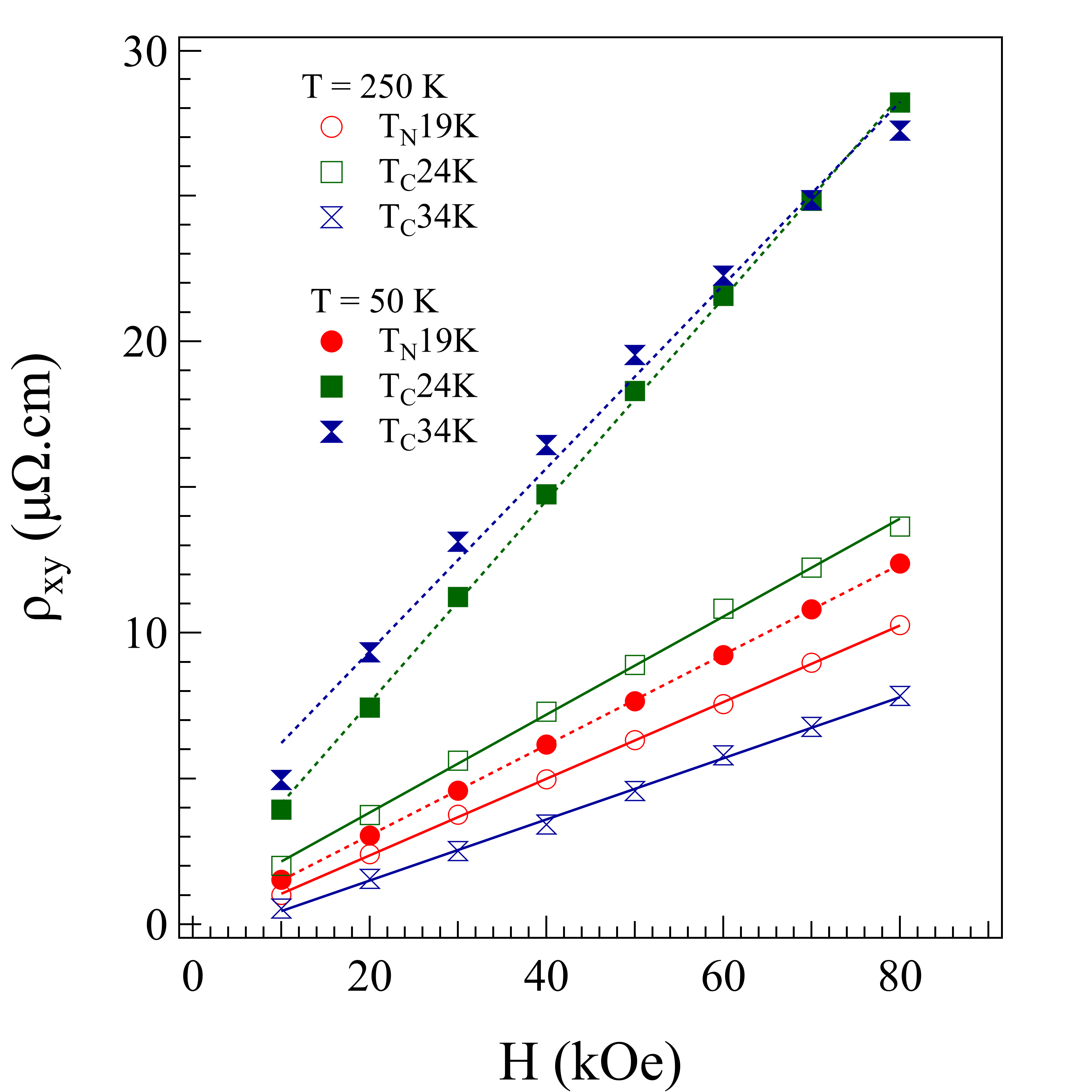}
\caption{(color online) Hall resistivity versus applied magnetic field for temperatures at 50\,K and 250\,K with the electrical current in the \textit{ab}-plane and the magnetic field along the \textit{c}-axis. }
\label{fig:Rxy}
\end{figure}

\begin{figure} \centering \includegraphics [width = 0.47\textwidth] {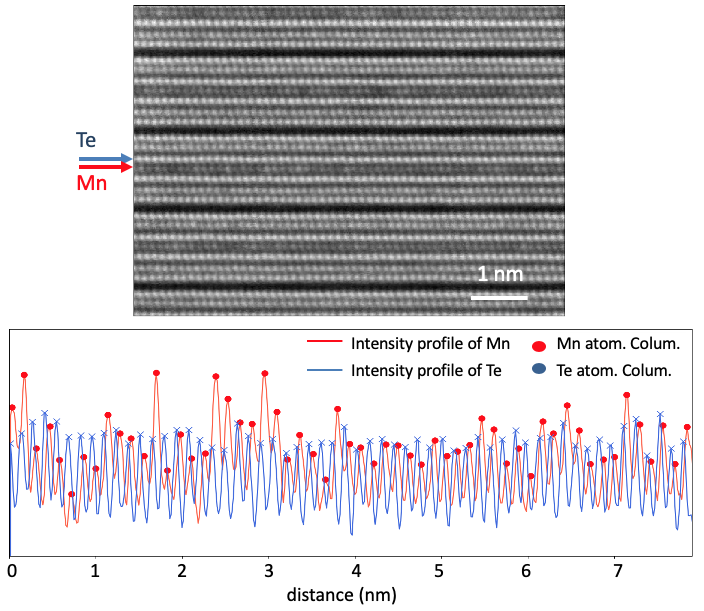}
\caption{(color online) (a) STEM-HAADF images of T$_C$24K. (b) Intensity variation in Mn and Te atomic columns. }
\label{fig:STEMProfile}
\end{figure}

\subsection{Transport properties}
Figure~\ref{fig:Rxx} shows the temperature and field dependence of the in-plane electrical resistivity, $\rho_{xx}$, measured in the temperature range 2~K~$\leq T \leq$~300~K with the electrical current flowing in the \textit{ab}-plane. All compounds show a metallic conducting behavior in the paramagnetic state.  $\rho_{xx}$  shows an abrupt enhancement cooling across $T_N$ for T$_N$19K. In contrast, a cusp sitting around $T_C$ was observed for T$_C$24K and T$_C$34K, which is similar to that observed for MnBi$_2$Te$_4$. All compounds show a negative magnetoresistance in magnetic fields up to 80\,kOe applied along the crystallographic \textit{c}-axis.

The Hall resistivity was measured at multiple temperatures in the temperature range 2~K~-~300~K. A linear field dependence is observed at all temperatures for all samples. Figure~\ref{fig:Rxy} shows the data collected at 50\,K and 250\,K. The Hall coefficient shows little temperature dependence below room temperature. With the assumption that a single band dominates the Hall signal, the coefficient at 250~K gives a carrier density of  2.8$\times$10$^{20}$cm$^{-3}$,  and 3.7$\times$10$^{20}$cm$^{-3}$, and 6.0$\times$10$^{20}$cm$^{-3}$ for T$_N$19K, T$_C$24K, and T$_C$34K, respectively.

\subsection{STEM}
Figure~\ref{fig:STEMProfile} shows the intensity variation in Mn atomic columns along (001) planes indicating the presence of atomic mixing. The intensity profiles shown in Fig. S7 b are obtained from the representative atomic planes of Te and Mn labeled in (a). Obviously, Mn atomic columns show a larger intensity variation than Te columns. This agrees with both our diffraction data and previous DFT calculations~\cite{du2020tuning} that there are significant amount of site mixing at the Mn site but not at the Te site.

\subsection{DFT}

\begin{figure} \centering \includegraphics [width = 0.48\textwidth] {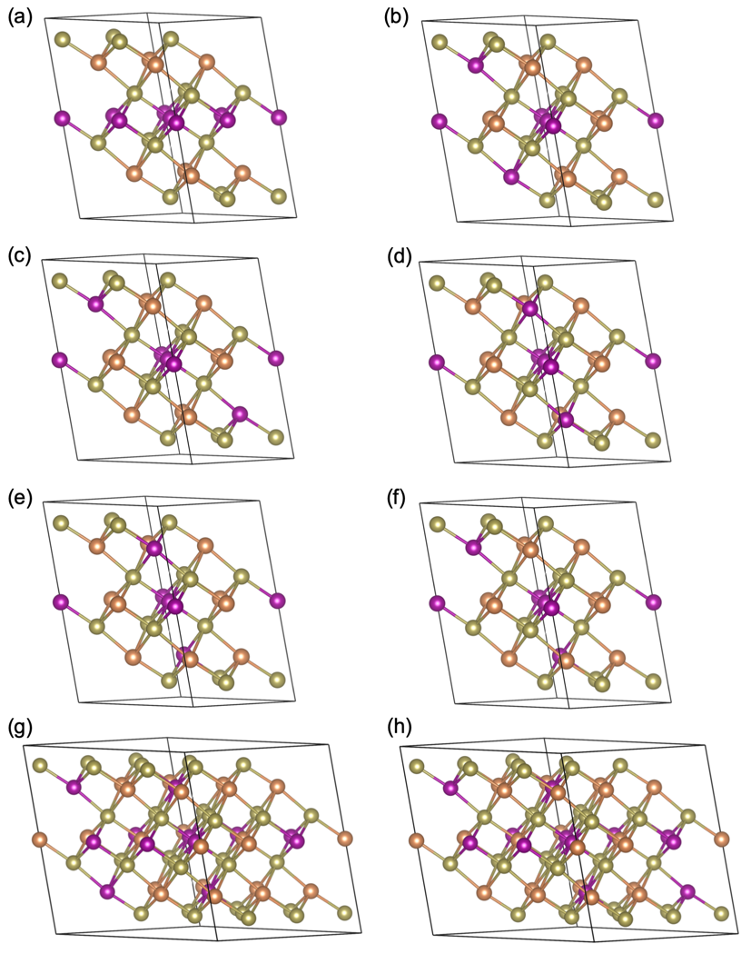}
\caption{(color online) MnSb$_2$Te$_4$ atomic configurations for half of the supercell of (a) (1$\times$1$\times$2)-AS0 without site mixing, (b) (2$\times$2$\times$2)-AS1, (c) (2$\times$2$\times$2)-AS2, (d) (2$\times$2$\times$2)-AS3, (e) (2$\times$2$\times$2)-AS4, (f) (2$\times$2$\times$2)-AS5, (g) (3$\times$3$\times$2)-AS1 and (h) (3$\times$3$\times$2)-AS2 with Mn (purple), Sb (orange) and Te (green) atoms. }
\label{fig:Config}
\end{figure}

\begin{figure} \centering \includegraphics [width = 0.48\textwidth] {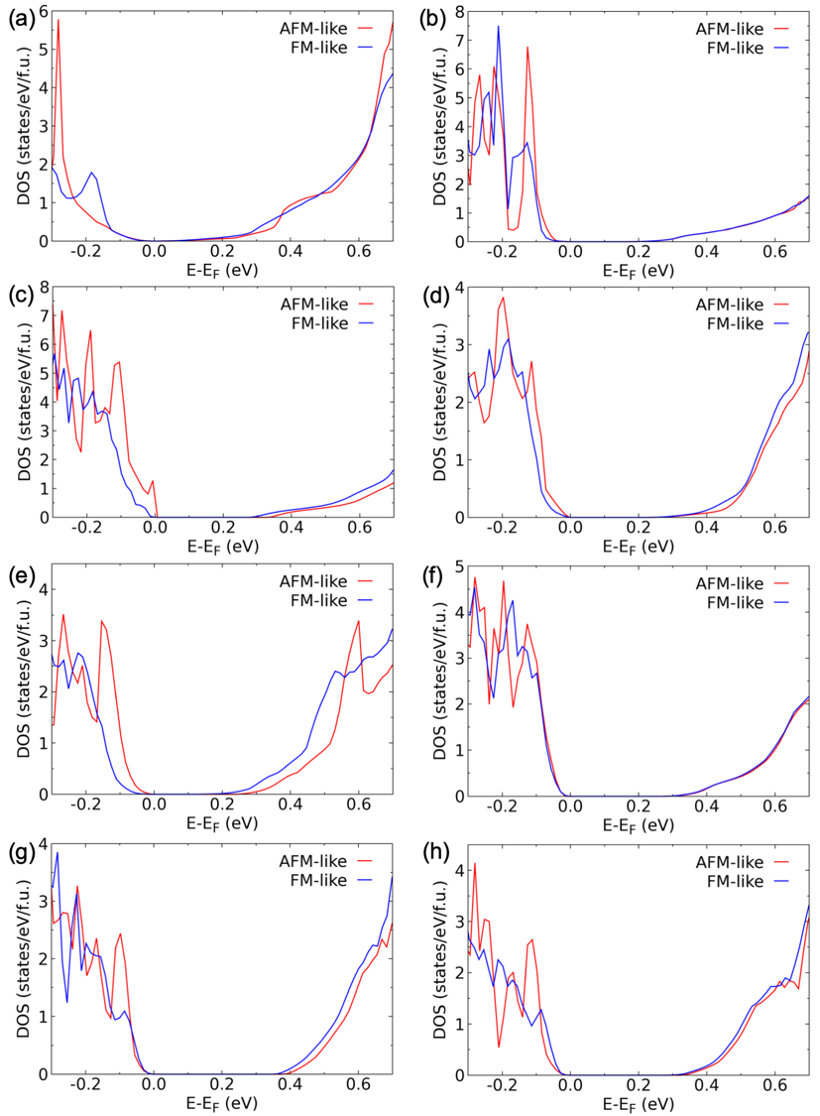}
\caption{(color online) Total density of states (DOS) near E$_F$ of the FM-like and AFM-like magnetic orders for (a) (1$\times$1$\times$2)-AS0 without site mixing, (b) (2$\times$2$\times$2)-AS1, (c) (2$\times$2$\times$2)-AS2, (d) (2$\times$2$\times$2)-AS3, (e) (2$\times$2$\times$2)-AS4, (f) (2$\times$2$\times$2)-AS5, (g) (3$\times$3$\times$2)-AS1 and (h) (3$\times$3$\times$2)-AS2. }
\label{fig:totalDOS}
\end{figure}

\begin{figure} \centering \includegraphics [width = 0.48\textwidth] {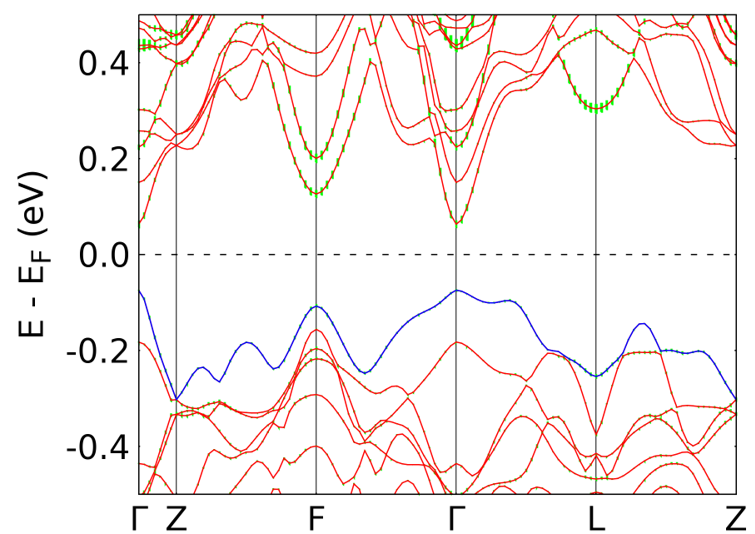}
\caption{(color online) PBE+U+SOC band structure with a FM alignment of magnetic moments on Mn ions at different sites. }
\label{fig:DFT2}
\end{figure}

Figure\,\ref{fig:Config} shows the atomic configurations for half of the supercell of MnSb$_2$Te$_4$ with different site-mixing configurations. Also shown are two  configurations in the larger (3$\times$3$\times$2) supercell with a smaller site-mixing ratio. Table\,\ref{tab:EnergyDiff} shows the energy difference between the fully relaxed FM-like and AFM-like magnetic order calculated in PBEsol+U and PBEsol+U+SOC~\cite{perdew2008restoring,dudarev1998electron}. All the supercell configurations with site-mixing prefer the FM-like magnetic order. Figure\,\ref{fig:totalDOS} shows the total density of states (DOS) near E$_F$ of the FM-like and AFM-like magnetic orders with different site-mixing configurations. Figure\,\ref{fig:DFT2} shows the band structure assuming all magnetic moments on Mn ions at different sites are aligned ferromagnetically. The band gap is reduced, but still no band inversion is expected.

\begin{table}[htbp]
\caption{Energy difference between the fully relaxed FM-like and AFM-like ($\bigtriangleup$E$_{FM-AFM}$) calculated in PBEsol+U and PBEsol+U+SOC for the configurations in the supercells of (1$\times$1$\times$2), (2$\times$2$\times$2) and (3$\times$3$\times$2). The (1$\times$1$\times$2)-AS0 is the ideal configuration without anti-site (AS) mixing. The other supercell AS configurations in the septuple layer are shown in Fig. ~\ref{fig:Config} and named numerically.}
\centering
%\begin{tabular}{c|c|c}
\begin{tabular}{|C{2.7cm}|C{2.7cm}|C{2.7cm}|}
  \hline
   \hline
   
                      & $\Delta$E$_{FM-AFM}$ &  $\Delta$E$_{FM-AFM}$  \\
Configuration &  (meV/f.u.) &   (meV/f.u.) \\
                      &  PBEsol+U &   PBEsol+U+SOC \\
  \hline
 (1$\times$1$\times$2)-AS0& +3.48 &  +3.55  \\
  \hline
 (2$\times$2$\times$2)-AS1   & -0.03  &  -0.02   \\
  \hline
   (2$\times$2$\times$2)-AS2& -0.84 &  -0.93  \\
  \hline
 (2$\times$2$\times$2)-AS3   & -0.31  &  -0.30   \\
  \hline
    (2$\times$2$\times$2)-AS4& -0.10 &  -0.11  \\
  \hline
 (2$\times$2$\times$2)-AS5   & -0.54  &  -0.70   \\
  \hline
    (3$\times$3$\times$2)-AS1& -0.15 &  -0.12  \\
  \hline
 (3$\times$3$\times$2)-AS2   & -0.12  &  -0.08   \\
  \hline
 \end{tabular}
\label{tab:EnergyDiff}
\end{table}

\subsection{STS}

\begin{figure} \centering \includegraphics [width = 0.48\textwidth] {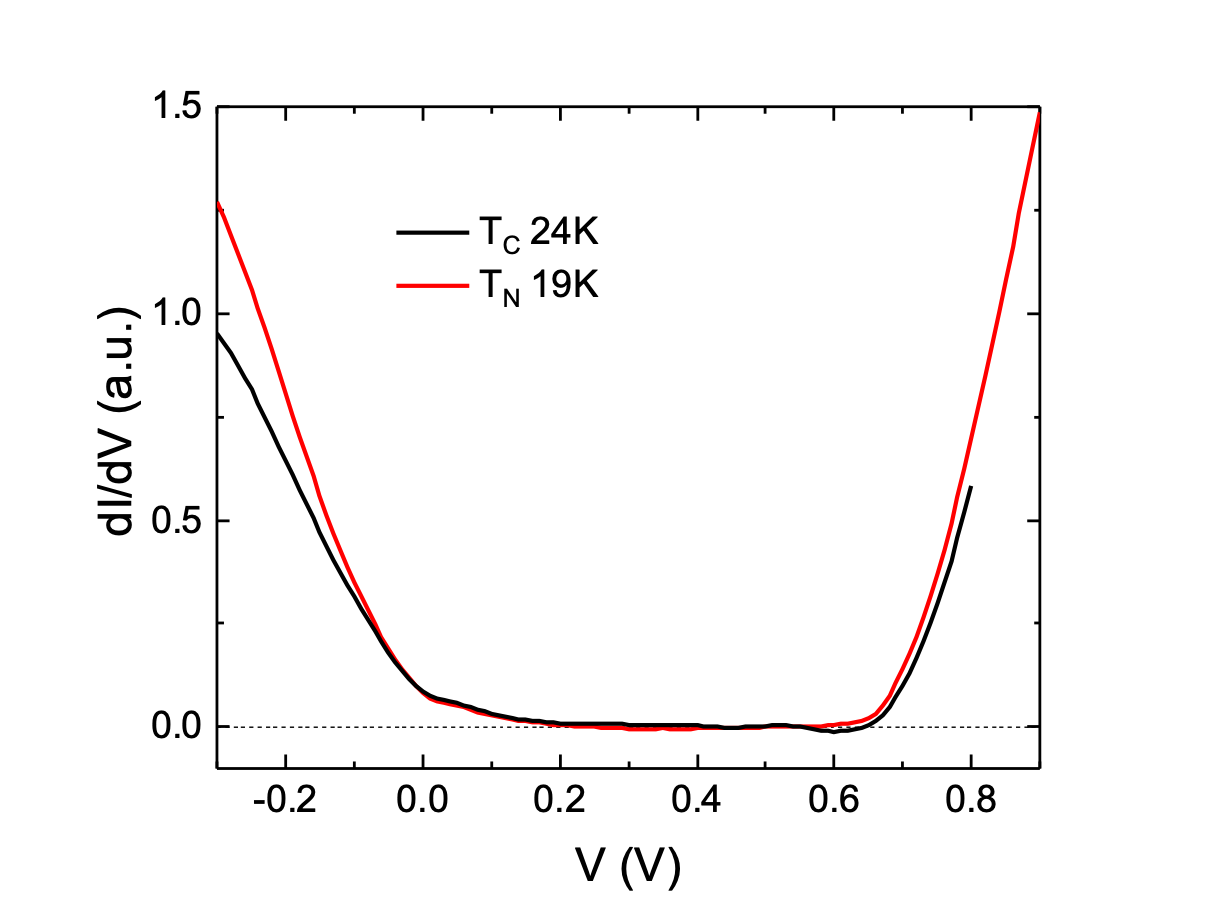}
\caption{(color online) Scanning tunneling spectra of T$_C$24K (black) and T$_N$19K (Red) samples measured at 4.5 K. The tunneling setpoint for T$_C$24K is –0.3 V, 0.4 nA, and that for T$_N$19K is –0.3 V, 0.5 nA. }
\label{fig:STS}
\end{figure}

To validate the electronic structure from DFT calculations, we measured scanning tunneling spectra (STS) of T$_N$19K and T$_C$24K at 4.5\,K. As shown in Fig.\,\ref{fig:STS}, both samples have a bulk band gap of about 0.4 eV, with valence band maximum at 0.2 eV above E$_F$ and conduction band minimum at 0.6 eV. The flat and diminished DOS within the bandgap indicates the absence of in-gap states and trivial topology of the band structures. The sizable band gap measured in the samples agrees with and can be used to justify the DFT-calculated band structure with the supercell approach as described above.

\subsection{references}
\bibliographystyle{apsrev4-1}
\bibliography{MST}

\end{document}